\documentclass{amsart}

\usepackage{a4wide} % wide format, as original submissions
\usepackage[T1]{fontenc} % correctly encode fonts in PDF
\usepackage[utf8]{inputenc} % correctly interpret encoding in TeX
\usepackage[foot]{amsaddr} % place author addresses on first page

\usepackage{mathpartir}
\usepackage{amsmath,amssymb}
\usepackage{times}
\usepackage{hyperref}
\usepackage{listings}

{\newtheorem{claim}{Claim}}

\makeatletter
\lstdefinestyle{andromedaStyle}{basicstyle=\ttfamily\lst@ifdisplaystyle\footnotesize\fi}
\makeatother

\makeatletter
\def\lst@visiblespace{ }
\makeatother
\title[Andromeda Proof Assistant]{Design and Implementation of the Andromeda Proof Assistant}

\author[A.~Bauer]{Andrej Bauer}
\address{Andrej Bauer \\ Faculty of mathematics and Physics (University of Ljubljana) \\ and Institute for mathematics Physics and Mechanics}

\email{Andrej.Bauer@andrej.com}
\thanks{This material is based upon work supported by the Air Force Office of Scientific Research, Air Force Materiel Command, USAF under Award No. FA9550-14-1-0096}

\author[G.~Gilbert]{Gaëtan Gilbert}
\address{Gaëtan Gilbert \\ École Normale Supérieure de Lyon \\ and INRIA (Nantes)}
\email{gaetan.gilbert@skyskimmer.net}

\author[P.~G.~Haselwarter]{Philipp G.~Haselwarter}
\address{Philipp G.~Haselwarter \\ Faculty of mathematics and Physics\\ University of Ljubljana}
\email{philipp@haselwarter.org}

\author[M.~Pretnar]{Matija Pretnar}
\address{Matija Pretnar \\ Faculty of mathematics and Physics\\ University of Ljubljana}
\email{matija@pretnar.info}

\author[C.~A.~Stone]{Christopher A.~Stone}
\address{Christopher A.~Stone \\ Harvey Mudd College}
\email{stone@cs.hmc.edu}

\newcommand{\G}{\Gamma} % a context
\newcommand{\D}{\Delta} % another context
\newcommand{\E}{\Xi} % yet another context

\renewcommand{\AA}{A} % a type
\newcommand{\BB}{B} % another type
\newcommand{\CC}{C} % another type
\newcommand{\DD}{D} % another type

\newcommand{\x}{x} % a term variable
\newcommand{\y}{y} % another term variable
\newcommand{\sss}{s} % another term
\newcommand{\ttt}{t} % another term
\newcommand{\uuu}{u} % another term
\newcommand{\vvv}{v} % another term

\newcommand{\ctxdom}[1]{\mathrm{dom}(#1)}

\newcommand{\rulename}[1]{\text{\textsc{#1}}}

\newcommand{\bnf}{\ \mathrel{{:}{:}{=}}\ }
\newcommand{\bnfor}{\ \mid\ \ }

\newcommand{\ctxempty}{\bullet} % empty context
\newcommand{\ctxextend}[3]{#1,\, #2\, {:}\, #3} % extended context

\newcommand{\subst}[3]{#1[#3/#2]} % substitution
\newcommand{\Type}{\mathsf{Type}} % The type of all types
\newcommand{\Prod}[2]{\mathop{\textstyle\prod_{(#1 {:} #2)}}} % dependent product

\newcommand{\lam}[3]{\lambda #1 {:} #2.{#3}\,.\,} % $\lambda$-abstraction
\newcommand{\app}[5]{#1\mathbin{@^{#2{:}#3.#4}} #5} % application

\newcommand{\JuEqual}[3]{\mathsf{Eq}_{#1}(#2,#3)} % Equality type
\newcommand{\juRefl}[1]{{\mathsf{refl}_{#1}}\ }    % Judgmental refl

\newcommand{\isctx}[1]{#1\ \mathsf{ctx}} % well formed context
\newcommand{\istype}[2]{#1 \vdash #2\ \mathsf{type}} % well formed type
\newcommand{\isterm}[3]{#1 \vdash #2 : #3} % well formed term

\newcommand{\istermI}[3]{#1 \vdash #2 : #3} % well formed term (NB: currently the same as \isterm!)

\newcommand{\eqterm}[4]{#1 \vdash #2 ~ \equiv ~ #3 : #4} % equal terms
\newcommand{\eqtype}[3]{#1 \vdash #2 \equiv #3}

\newcommand{\Nat}{\mathsf{Nat}} % the type of natural numbers
\newcommand{\Bool}{\mathsf{Bool}} % the type of booleans

\newcommand{\ctxrestrict}[2]{#1{\restriction}_{#2}}
\newcommand{\ctxjoin}[2]{#1 \bowtie #2}

\newcommand{\naturalTy}[1]{\mathcal{N}(#1)}

\newcommand{\judgTy}{\mathtt{judgment}}

\newcommand{\cType}{\mathtt{Type}}
\newcommand{\cProd}[3]{\Pi (#1 {:} #2),\, #3} % can we have ttfont Pi?
\newcommand{\capp}[2]{#1 \, #2}
\newcommand{\clam}[2]{\lambda (#1 {:} #2),\,} % can we have ttfont lambda?
\newcommand{\cEq}[2]{#1 \equiv #2} % can we have ttfont \equiv?
\newcommand{\crefl}[1]{\mathtt{refl}\,#1}
\newcommand{\cassume}[2]{\mathtt{assume}\;#1\,{:}\,#2\;\mathtt{in}\;}
\newcommand{\csubst}[3]{#1\,\mathtt{where}\,#2 = #3}
\newcommand{\cascribe}[2]{#1:#2}
\newcommand{\cnatural}[1]{\mathtt{natural}\,#1}

\newcommand{\ccontext}[1]{\mathtt{context}\,#1}
\newcommand{\cdynamic}[2]{\mathtt{dynamic}\;#1 = #2}
\newcommand{\cnow}[2]{\mathtt{now}\;#1 = #2\;\mathtt{in}\;}
\newcommand{\coccurs}[2]{\mathtt{occurs}\,#1\,#2}
\newcommand{\cyield}[1]{\mathtt{yield}\,#1}

\newcommand{\jpatt}[2]{\vdash #1 : #2}
\newcommand{\pvar}[1]{\mathtt{?}#1}
\newcommand{\patom}[1]{\text{\lstinline{_atom}}\;#1}
\newcommand{\pconst}[1]{\text{\lstinline{_constant}}\;#1} 
\newcommand{\opEqual}[2]{\mathtt{equal}\,#1\,#2}
\newcommand{\opAsProd}[1]{\mathtt{as{\char95}prod}\,#1}
\newcommand{\opAsEq}[1]{\mathtt{as{\char95}eq}\,#1}
\newcommand{\opCoerce}[2]{\mathtt{coerce}\,#1\,#2}
\newcommand{\opCoerceFun}[1]{\mathtt{coerce{\char95}fun}\,#1}

\begin{document}

\maketitle

\begin{abstract} Andromeda is an LCF-style proof assistant where the user
  builds derivable judgments by writing code in a meta-level programming
  language AML. The only trusted component of Andromeda is a minimalist nucleus
  (an implementation of the inference rules of an object-level type theory),
  which controls construction and decomposition of type-theoretic judgments.

  Since the nucleus does not perform complex tasks like equality checking beyond
  syntactic equality, this responsibility is delegated to the user, who
  implements one or more equality checking procedures in the meta-language. The
  AML interpreter requests witnesses of equality from user code using the
  mechanism of algebraic operations and handlers. Dynamic checks in the nucleus
  guarantee that no invalid object-level derivations can be constructed.
  %even if the AML code (or interpreter) is untrusted.

  To demonstrate the flexibility of this system structure, we implemented
  a nucleus consisting of dependent type theory with equality reflection.  Equality
  reflection provides a very high level of expressiveness, as it allows the user
  to add new judgmental equalities, but it also destroys desirable
  meta-theoretic properties of type theory (such as decidability and strong
  normalization).

  The power of effects and handlers in AML is demonstrated by a standard library
  that provides default algorithms for equality checking, computation of normal
  forms, and implicit argument filling.  Users can extend these new algorithms
  by providing local ``hints'' or by completely replacing these algorithms for
  particular developments.  We demonstrate the resulting system by showing how
  to axiomatize and compute with natural numbers, by axiomatizing the untyped
  $\lambda$-calculus, and by implementing a simple automated system for managing
  a universe of types.  \end{abstract}

\section{Introduction}\label{sec:introduction}

A type theory can be interesting and very useful, yet lack metatheoretic
properties (e.g., decidability) that permit a straightforward implementation.
In fact, the more flexible and expressive the theory, the less likely these
properties will hold. Nevertheless, even very expressive type theories deserve
automated support in the form of proof assistants. The question is how a useful
proof assistant can make minimal demands on the properties
of the underlying object language. In this paper, we describe the structure of one
such system.

\emph{Andromeda} is an LCF-style proof assistant~\cite{lcf} in which (derivable)
judgments are the fundamental data of the system. These judgments are opaque
except within a tiny, trusted nucleus that implements rules of the
underlying type theory (to construct new judgments from old) and also implements valid
inversion principles (to decompose judgments into sub-judgments).  The
untrusted remainder of the hard-coded system is a small interpreter for AML, an
ML-like meta-language~\cite{standard-ml} extended with algebraic operations and
handlers~\cite{algebraic-effects}.

The AML interpreter builds and decomposes judgments by making (dynamically
checked) requests of the trusted nucleus. When these requests would fail (e.g., because
a function is being applied to an argument, and in violation of the appropriate
typing rule the domain type of the function
is not syntactically identical to the type of the argument), the interpreter
triggers a suitable algebraic operation to request additional information (e.g.,
evidence of equality between the mismatched types) from the user.

User-level AML code directs the construction of judgments, and consists
of computations that construct and pattern-match judgment values
and user-level handlers that intercept and respond to algebraic
operations triggered during judgment construction. The consequence
of this design is that most proof-assistant functionality---including equality checking,
normalization, unification, and proof tactics---is handled at the user
level. Effects and handlers allow default implementations (necessarily
incomplete for an undecidable object language) that can be overridden using
nested handlers, providing specialized algorithms for specific trouble spots.

\smallskip

The specific expressive object language is largely independent of this system
design, but some readers may find our chosen type theory independently
interesting.  The type theory currently implemented in Andromeda is dependent
type theory with \emph{equality reflection}, the principle that propositionally
equal terms are judgmentally equal:
\begin{equation*}
  \infer
  {\isterm{\G}{\uuu}{\JuEqual{\AA}{\sss}{\ttt}}}
  {\eqterm{\G}{\sss}{\ttt}{\AA}}
\end{equation*}

From a mathematical point of view, equality reflection is appealing and natural,
as it makes equality in type theory behave like ordinary equality in
mathematics. (In Coq, for example, the types ``vector of length $0+n$'' and
``vector of length $n$'' are equal because $0+n$ and $n$ are judgmentally equal,
but a ``vector of length $n+0$'' requires an explicit coercion to be used
as a ``vector of length $n$'' because $n+0$ is only propositionally
equal to $n$.) From the perspective of homotopy type theory,
equality reflection is suitable for ``set-level'' mathematics, i.e., those
mathematical structures that do not exhibit any higher homotopical phenomena.
Among these are substantial parts of algebra, analysis, and logic, including
many aspects of meta-theory of type theory.

Building equality reflection into a proof assistant has practical advantages.
First, equality reflection lets users axiomatize
type-theoretic constructions such as natural numbers with \emph{judgmental} equalities,
meaning that we can implement a smaller trusted core type theory
with a wider variety of possible user extensions.
Second, applications of equality reflection are not recorded in the conclusion,
and omitting the explicit equality eliminators keeps
terms smaller and simpler.
%The explicit equality eliminators in intensional type theory
%can obstruct and complicate matters without
%adding any insight.

The proof assistant NuPRL~\cite{nuprl} validated equality reflection by
implementing so-called \emph{computational} type theory, a specific
interpretation of type theory akin to realizability models.
More recently, however, equality reflection has fallen into disrepute
among computer scientists and computationally minded mathematicians.
It causes the loss of useful meta-theoretic properties such as
strong normalization of terms and decidability of type
checking~\cite{hofmann97:_exten}, the cornerstones of modern proof
assistants like Coq~\cite{coq}, Agda~\cite{agda} and Lean~\cite{lean}.
Even the property
``if an application of a lambda abstraction to an argument is well typed,
then its $\beta$-reduct is well typed'' may not hold if the
user assumes nonstandard type equalities.

Nevertheless, the use of effects and handlers allows Andromeda to take advantage
of equality reflection and to deal with its negative consequences gracefully.

\subparagraph*{Contributions}

The present paper should be read as a progress report on the development of
Andromeda; the system and the underlying type theory may evolve as we gain more
experience and consider a wider variety of applications. We focus
on the following points of interest:

\begin{itemize}
\item the goals of Andromeda and the structure of the system (\S\ref{sec:system-design});
\item the impact of equality reflection on both the design of the type-theoretic nucleus
  and the details of its implementation (\S\ref{sec:nucleus},
  Appendix~\ref{sec:rules-type-theory});
\item features of the meta-language that allow a variety of
  proof-development techniques to be implemented at the user level (\S\ref{sec:meta-language});
\item a discussion of the soundness of the system (\S\ref{sec:soundness-andromeda});
\item a prototype standard library that provides user-extensible equality checking and
  implicit-argument filling (\S\ref{sec:stand-libr});
\item axiomatization of additional type-theoretic structures (dependent sums, natural
  numbers, untyped $\lambda$-calculus, and universes), \emph{with} the desired judgmental
  equalities and support for automation (\S\ref{sec:examples}).
\end{itemize}

Andromeda is free software, available at \url{http://www.andromeda-prover.org/}.
Contributions, questions, and requests are most welcome.

%%% Local Variables:
%%% mode: latex
%%% TeX-master: "main"
%%% End:

\section{An overview of Andromeda}
\label{sec:system-design}

Andromeda follows design principles that are similar to those of other proof assistants:
\begin{itemize}
\item
The system should \emph{work well in the common case}. Equality reflection
affords many possibilities for complicating one's life, but we expect most applications to be very
reasonable. If the user introduces new computation and extensionality rules that
play nicely with the existing ones, the system should work smoothly.
Nevertheless, less common scenarios should still work, possibly with
more effort on the part of the user.

\item
The user cannot be expected to write down explicit typing annotations on all terms, or
hold in their head various bureaucratic matters, such as the typing contexts. Therefore,
the \emph{system should take care of low-level details}.

\item
There should be a \emph{clearly delineated nucleus} that is the only part of the
implementation that the user has to trust\footnote{Except for trusting the OCaml compiler, the operating system,
  the hardware, and the absence of malicious cosmic rays.}  in
order to believe that Andromeda never produces an invalid judgment. The nucleus should be
as small as possible and its functionality should implement type theory in the most
straightforward way possible.

\item
A consequence of this minimalism is that the system should be \emph{user
  extensible}, so that additional functionality can be introduced without breaking trust.
\end{itemize}

Andromeda is implemented in the tradition of Robin Milner’s
Logic for Computable Functions (LCF)~\cite{lcf}.
The current implementation, in OCaml~\cite{ocaml}, consists of around 9500 lines of source code, of which the
nucleus comprises 1900 lines. These are very low numbers that clearly classify Andromeda
as a prototype. However, we do not expect the nucleus to grow significantly.

%\iffalse

The core of Andromeda is the trusted nucleus that directly implements inference
rules and inversion rules for dependent type theory with equality
reflection. By design, it is the only part of the system that can create and
manipulate type-theoretic judgments. The nucleus is
small and simple, as it does not perform \emph{any} proof search, unification,
equality checking, or normalization. (It cannot, since equality checking is in
general undecidable and there is no reasonable notion of normal
form~\cite{hofmann97:_exten}.) Whenever evidence of equality is needed as
a premise to an inference rule, it must be provided to the nucleus explicitly.

The user interacts with the system by writing code in the \emph{Andromeda meta-language
  (AML)}, a general-purpose programming language in the style of~ML\@. AML exposes the
nucleus datatype~$\judgTy$ as an abstract datatype of its own. Because judgments may only
be constructed by the nucleus, neither the OCaml implementation of AML nor any user code
written in AML need be trusted. AML handles only trivial syntactic equality checks. All
other evidence of equality is obtained from user-level code, through the mechanism of
algebraic operations and handlers~\cite{handlers,eff-lang} (\S\ref{sec:operations-handlers}).

Users are free to organize their AML code in any way they see fit. In most
cases they would likely want a good axiomatization of standard type constructors
(dependent sums, inductive types, universes, etc.), equality checking
algorithms that work well in the common cases, and conveniences such as
resolution of implicit arguments. These ought to be provided by a standard
library (\S\ref{sec:stand-libr}).  In principle,
there may be several such libraries, or even several equality checking
algorithms in a single library. The handlers mechanism allows flexible and local
uses of several different equality checking algorithms.

AML is statically typed---and this caught many silly errors while we were
coding a standard library---but the AML type system is unrelated to
the soundness of the system. Bugs in AML code either
prevent code from constructing the desired judgments or
construct an unintended judgment, but the abstract type of
judgments and run-time checks in the nucleus ensure that only derivable
judgments are ever constructed. Any other memory-safe metalanguage (e.g., one modeled
on Python or Scheme) would be equally sound, if less robust.

\paragraph*{Andromeda in action}

Before looking at the three constituent parts of Andromeda in more detail, we provide a small worked example. At this point we cannot explain all the
technical details, so we focus on emphasizing the important points and showcasing what
Andromeda can do.

We begin by declaring some constants that Andromeda adds to the ambient signature:
\begin{lstlisting}
constant A : Type
constant a : A
constant b : A
constant P : A → Type
constant v : P a
\end{lstlisting}
Andromeda manipulates \emph{only} judgments. Thus the above declaration binds
the AML variable \lstinline{a} to the nucleus judgment \lstinline{⊢ a : A}, not
to a bare symbol (and similarly for \lstinline{b}, \lstinline{v}, \lstinline{A}
and \lstinline{P}). Nevertheless, it is often convenient to think of a judgment as ``a
term with a given type, possibly depending on some hypotheses''.

Let us first show that the type of transport is inhabited:
\begin{lstlisting}
∏ (x y : A), x ≡ y → P x → P y
\end{lstlisting}
In intensional type theory we would use a $J$ eliminator, but here we should be able to use the
curried term \lstinline{λ x y ξ u, u}. Indeed, \lstinline{u} may be converted from \lstinline{P x}
to \lstinline{P y} because these are equal types by an application of the congruence rule
for applications and equality reflection of \lstinline{ξ : x ≡ y}. The Andromeda standard
library, which is implemented at the user-level, does all this for us if we tell it to use
\lstinline{ξ} as an \emph{equality hint} while checking that \lstinline{u} has type \lstinline{P y}:
\begin{lstlisting}
λ (x y : A) (ξ : x ≡ y) (u : P x), (now hints = add_hint ξ in (u : P y))
\end{lstlisting}
The above is \emph{not} a proof term but an AML computation that generates a judgment.
In particular, AML immediately evaluates the command inside the~\lstinline{λ}.
While doing so it will find it needs a witness for the equality
between \lstinline{P x} (the type of \lstinline{u}) and \lstinline{P y} (the
type ascribed to \lstinline{u}); it requests one using the operations and handler mechanism.
The standard library handles this, employing an equality checking algorithm that
eventually uses the hint \lstinline{ξ} to equate \lstinline{x} and \lstinline{y}, and passing back the
requested evidence to AML.
Then, AML asks the nuclues to apply equality reflection to obtain the judgmental equality
of \lstinline{P x} and \lstinline{P y} (at which point the equality witness provided by the
standard library is discarded) and to apply conversion. The interaction betwen AML and the nuclues
proceeds in this fashion until the judgment witnessing transport is constructed.

If we need to write \lstinline{add_hint} to guide the type checker, one might ask why this
is better than the intentional approach of applying a \lstinline{J} eliminator with
\lstinline{ξ} to coerce \lstinline{u} from \lstinline{P x} to \lstinline{P y}. A single
\lstinline{add_hint} is like the incorporation of a computation rule that can handle an
arbitrarily complex development, whereas \lstinline{J} is like a single application of a
computation rule that has to be repeated at every point where the rule is to be applied.

The other benefit, apparent even here, is that without \lstinline{J} the proof term is smaller.
In the end, the judgment built by the nucleus is the expected one:
\begin{lstlisting}
⊢ λ (x : A) (y : A) (_ : x ≡ y) (u : P x), u
  : Π (x : A) (y : A), x ≡ y → P x → P y
\end{lstlisting}

Although \lstinline{ξ} does not appear explicitly in the conclusion,
Andromeda is aware it was used. This tracking process becomes
apparent if we \emph{temporarily} hypothesize an equality \lstinline{a ≡ b} and use it as a hint while constructing a judgment that \lstinline{v} above has type \lstinline{P b},
\begin{lstlisting}
assume ζ : a ≡ b in
  now hints = add_hint ζ in (v : P b)
\end{lstlisting}
This AML expression causes the nucleus to build the \emph{hypothetical} judgment:
\begin{lstlisting}
ζ₀ : a ≡ b ⊢ v : P b
\end{lstlisting}
The \lstinline{assume} construct generated a fresh variable \lstinline{ζ₀} of type
\lstinline{a ≡ b} and bound the AML variable \lstinline{ζ} to the judgment
\lstinline{ζ₀ : a ≡ b ⊢ ζ₀ : a ≡ b}.
Because \lstinline{ζ₀} was used to convert \lstinline{P a} to \lstinline{P v}, the
nucleus produced a judgment that depends on it.

The AML interpreter communicates with user-level code by invoking operations to be handled, but user-level operations are useful as well.
Let us define a simple \lstinline{auto} tactic for automatically inhabiting simple types.
We first need an AML function, called \lstinline{derive}, which attempts to inhabit a
given type from the currently available hypotheses by performing a recursive search. This
takes about 40 lines of uneventful code, shown in Appendix~\ref{sec:auto-tactic}.
Then we declare a new operation that takes no arguments and yields judgments,
\begin{lstlisting}
operation auto : judgment
\end{lstlisting}
and install a global handler that handles it:
\begin{lstlisting}
handle
| auto : ?Surr ⇒
    match Surr with
    | Some ?T ⇒ derive T
    | None    ⇒ failure
    end
end
\end{lstlisting}
When the handler intercepts the operation \lstinline{auto}, the surroundings
of the occurrence of \lstinline{auto} may or may not have indicated an expected result type~\lstinline{T}. If it does, the handler
calls \lstinline{derive T} to inhabit the type, otherwise it fails by triggering the
operation \lstinline{failure} (also defined in the appendix) because it has no information
on what type to inhabit.

Now we can use \lstinline{auto} to inhabit types. For example,
\begin{lstlisting}
λ (X : Type), (auto : X → X)
\end{lstlisting}
constructs the judgment
\begin{lstlisting}
⊢ λ (X : Type) (x : X), x
  : Π (X : Type), X → X
\end{lstlisting}
Given types \lstinline{A}, \lstinline{B}, and \lstinline{C}, the computation
\begin{lstlisting}
auto : (A → B → C) → (A → B) → (A → C)
\end{lstlisting}
results in the judgment
\begin{lstlisting}
⊢ λ (x : A → B → C) (x0 : A → B) (x1 : A), x x1 (x0 x1)
  : (A → B → C) → (A → B) → A → C
\end{lstlisting}

The Andromeda standard library (\S\ref{sec:stand-libr}) takes full advantage of
operations and handlers to produce equality proofs and coercions, with default
implementations that users can override with local handlers when the standard heuristics
fail.

%%% Local Variables:
%%% mode: latex
%%% TeX-master: "main"
%%% End:

\section{The nucleus}
\label{sec:nucleus}

The nucleus is the part of the system that implements the object-level type theory. Its
functionality includes the following:
\begin{itemize}
\item formation and decomposition of term and type judgments,
\item construction of equality judgments,
\item substitution and syntactic equality checking,
\item pretty-printing of judgments and export to JSON.
\end{itemize}
Before discussing some of these features we take a closer look at the type theory implemented by
Andromeda, and engineering issues that it raises.

\subsection{Type theory with equality reflection}
\label{sec:type-theory-with}

The Andromeda nucleus implements an extensional Martin-Löf type
theory~\cite{itt,hofmann97:_exten} with dependent products~$\Prod{\x}{\AA}{\BB}$
and equality types $\JuEqual{\AA}{\sss}{\ttt}$. Complete rules
are provided in Appendix~\ref{sec:rules-type-theory}.
Fundamentally, the system is not too far removed from the more common
intensional Martin-Löf type theory, but instead of a $J$ eliminator
for equality types, we have equality reflection and uniqueness of equality terms:
\begin{mathpar}
  \infer[\rulename{eq-reflection}]
  {\isterm{\G}{\uuu}{\JuEqual{\AA}{\sss}{\ttt}}}
  {\eqterm{\G}{\sss}{\ttt}{\AA}}

  \infer[\rulename{eq-eta}]
  {\isterm{\G}{\ttt}{\JuEqual{\AA}{\sss}{\uuu}} \\
    \isterm{\G}{\vvv}{\JuEqual{\AA}{\sss}{\uuu}}
  }
  {\eqterm{\G}{\ttt}{\vvv}{\JuEqual{\AA}{\sss}{\uuu}}}
\end{mathpar}
The $J$ eliminator can easily be derived from these rules, but direct use of equality
reflection is generally simpler. Streicher's $K$~eliminator and uniqueness of
identity proofs~\cite{streicher} are also derivable in this setting.

Equality reflection invalidates some common structural rules and inversion
principles, so we make further small changes to the type theory to compensate.
First, it is usual for products to satisfy an injectivity property, i.e., if
$\Prod{x}{\AA_1}{\AA_2}$ and $\Prod{x}{\BB_1}{\BB_2}$ are equal then $\AA_1$ equals
$\BB_1$ and $\AA_2$ equals $\BB_2$. But in our type theory injectivity fails because under
the assumption
\begin{equation}
  \label{eq:cantor-baire}
  p : \JuEqual{\Type}{(\Nat\to \Nat)}{(\Nat\to \Bool)}
\end{equation}
$\Nat \to \Nat$ and $\Nat \to \Bool$ are equal by reflection, \emph{without} equality of
$\Nat$ and $\Bool$.\footnote{The assumption that the Cantor space and the Baire space are
  equal may seem odd, but it is consistent. For instance, in classical set theory and in
  the effective topos the two are isomorphic, and with a little work we can arrange them to
  be equal.}
This may seem a very technical point, but usually one relies on injectivity to prove that
$\beta$-reductions preserve types. Indeed, under assumption~\eqref{eq:cantor-baire} the
identity function on $\Nat$ also has type $\Nat \to \Bool$, and hence by applying it
to~$0$ and $\beta$-reducing, we can show that~$0$ has type $\Bool$, even though $\Nat$ and
$\Bool$ are not equal.

Andromeda's solution, following~\cite{hofmann97:_exten}, is to add explicit
typing annotations that can typically be omitted in intentional type theories.
A $\lambda$-abstraction $\lam{\x}{\AA}{\BB} \ttt$ is annotated not only with the
domain $\AA$ of the bound variable but also with the type~$\BB$ of the
body~$\ttt$, and an application $\app{\sss}{\x}{\AA}{\BB}{\ttt}$ is similarly
annotated with the type of the function being applied. These annotations ensure
that terms have unique types up to equality: working again under the assumption~\eqref{eq:cantor-baire}, we can apply the identity function
at type $\Nat \to \Nat$ to get
$\app{(\lam{\x}{\Nat}{\Nat}{\x})}{\x}{\Nat}{\Nat}{0}$ of type $\Nat$, or
at $\Nat \to \Bool$ to get
$\app{(\lam{\x}{\Nat}{\Nat}{\x})}{\x}{\Nat}{\Bool}{0}$ of type $\Bool$.
Crucially, the typing annotations now prevent the latter term from $\beta$-reducing
to~$0$, as the $\beta$-rule requires that the function and the application match:
\begin{mathpar}
  \infer[\rulename{prod-beta}]
  {
    \isterm{\ctxextend{\G}{\x}{\AA}}{\sss}{\BB}\\
    \isterm{\G}{\ttt}{\AA}}
  {\eqterm{\G}{\app{(\lam{\x}{\AA}{\BB}{\sss})}{\x}{\AA}{\BB}{\ttt}}
              {\subst{\sss}{\x}{\ttt}}
              {\subst{\BB}{\x}{\ttt}}}
\end{mathpar}

Another principle that fails in the presence of equality reflection is \emph{strengthening},
which says that we may safely remove from the context any hypothesis that is not
explicitly mentioned in the conclusion of a judgment. Indeed,
\begin{equation*}
  \isterm{p : \JuEqual{\Type}{\Nat\to \Nat}{\Nat\to \Bool}}
         {\app{(\lam{\x}{\Nat}{\Nat}{\x})}{\x}{\Nat}{\Bool}{0}}
         {\Bool}
\end{equation*}
becomes invalid if we remove~$p$, even though there is no explicit use of~$p$ in the conclusion.
For similar reasons \emph{exchange} is not valid: given types $X$ and $Y$, the context
\begin{equation*}
    x : X,
    p : \JuEqual{\Type}{X}{Y},
    q : \JuEqual{Y}{x}{x}
\end{equation*}
becomes invalid if we exchange the order of $p$ and $q$, even though their types do not
refer to each other. The loss of strengthening and exchange is inconvenient; we discuss
an implementation-level solution in \S\ref{sec:implementation-of-type-theory}.

Perhaps the biggest difference between Andromeda and standard type theory is that
we currently postulate a single universe $\Type$ and the rule that makes
$\Type$ an element of itself:
\begin{mathpar}
  \infer[\rulename{ty-type}]
  {\isctx{\G}
  }
  {\isterm{\G}{\Type}{\Type}}
\end{mathpar}
From a logical point of view this is an inconsistent assumption, as Girard's paradox
implies that every type is inhabited~\cite{girard1972interpretation}. From an engineering
point of view, however, $\Type : \Type$ is very useful. For Andromeda implementers, it allows a
much simpler implementation strategy with fewer different judgment forms. For Andromeda
users, it allows postponing the complexities of dealing with type universes and universe
levels, and instead focus on other aspects of derivations. (For the same reason, both Coq
and Agda allow the assumption $\Type : \Type$ as an option.) Nevertheless, although users
are unlikely to stumble into inconsistencies by accident, we ultimately
want a sound foundation, and plan to remove $\rulename{ty-type}$, as discussed
in \S\ref{sec:remove-type-in-type}.

\subsection{Implementation of type theory}
\label{sec:implementation-of-type-theory}

The type theory implemented in the nucleus differs from the one presented in several ways.
The changes are inessential from a theoretical point of view, but have significant
practical impact. We describe them in this section.

\paragraph*{Signatures}

In Andromeda the user extends the type theory by postulating constants, i.e., they work in
type theory over a \emph{signature}. In this respect Andromeda is much like other proof
assistants that allow the user to state axioms and postulates. The signature is controlled
by the nucleus through an abstract datatype whose interface is very simple: there is an
empty signature, and a signature may be extended with a new constant of a given
\emph{closed} type (which may refer to the previously declared constants). Because
signatures are ever increasing, judgments derived over a signature remain valid when the
signature changes.

\paragraph*{Inversion principles and natural types}

The nucleus implements inversion principles for deconstruction of judgments into sub-judgments;
these are exposed in AML through pattern
matching, cf.~\S\ref{sec:datatype-judgments}. For example, an application
$\isterm{\G}{\app{\sss}{\x}{\AA}{\BB}{\ttt}}{\CC}$ can be decomposed into
$\isterm{\G}{\sss}{\Prod{\x}{\AA}{\BB}}$, $\isterm{\G}{\ttt}{\AA}$ and $\istype{\G}{\CC}$.
The type-theoretic justification for this operation is an \emph{inversion principle}:
if $\isterm{\G}{\app{\sss}{\x}{\AA}{\BB}{\ttt}}{\CC}$ is derivable then so are
$\isterm{\G}{\sss}{\Prod{\x}{\AA}{\BB}}$, $\isterm{\G}{\ttt}{\AA}$ and $\istype{\G}{\CC}$. Proving the
principle is not hard but neither is it a complete triviality, because the application
could have been formed with the use of type conversions. Similar inversion principles hold
for other term and type formers.

If we decompose an application, as above, and put it back together using the application
formation rule, we get the judgment
$\isterm{\G}{\app{\sss}{\x}{\AA}{\BB}{\ttt}}{\subst{\BB}{\x}{\ttt}}$. The application has
received its ``natural type'', which may not be the original type~$\CC$. Nevertheless, the
types are equal by \emph{uniqueness of typing}:
\begin{quote}
  \emph{If $\isterm{\G}{\ttt}{\AA_1}$ and $\isterm{\G}{\ttt}{\AA_2}$ then
    $\eqtype{\G}{\AA_1}{\AA_2}$.}
\end{quote}
The nucleus provides evidence of uniqueness of typing by generating, given
$\isterm{\G}{\ttt}{\AA}$, a witness for equality of $\AA$ and the \emph{natural type}
$\naturalTy{\ttt}$ of $\ttt$, which is read off the typing annotations:
\begin{align*}
  \naturalTy{\Type} &= \Type &
  \naturalTy{\Prod{\x}{\AA} \BB} &= \Type \\
  \naturalTy{\JuEqual{\AA}{\sss}{\ttt}} &= \Type &
  \naturalTy{\lam{\x}{\AA}{\BB} \ttt} &= \Prod{\x}{\AA}{\BB} \\
  \naturalTy{\app{\sss}{\x}{\AA}{\BB}{\ttt}} &= \subst{\BB}{\x}{\ttt} &
  \naturalTy{\juRefl{\AA} \ttt} &= \JuEqual{\AA}{\ttt}{\ttt}
\end{align*}
The natural types of variables and constants are read off the context and the signature,
respectively. Note that the natural type is the one we get if we deconstruct a term
judgment and construct it back again. In the standard library the equality of the original
type and the natural type is needed in several places during equality checking.

\paragraph*{Assumption sets} \label{sec:assumption-sets}

The nucleus is responsible for decomposing judgments into their component parts,
a facility used by pattern matching in AML. For example, we can combine
\begin{equation*}
\isterm{f : \Nat \to \Nat}{f}{\Nat \to \Nat}
\qquad\text{and}\qquad
\isterm{\x : \Nat}{\x}{\Nat}
\end{equation*}
(using weakening and application) to get
\begin{equation*}
\isterm
  {f:\Nat \to {\Nat},\, \x:\Nat}
  {\app{f}{\_}{\Nat}{\Nat}{x}}
  {\Nat},
\end{equation*}
But if we naively pattern-match on this application to get the function part and the
argument part (as judgments), we would get the constituents in weakened form
\begin{equation*}
\isterm{f : \Nat \to \Nat,\ \x:\Nat}{f}{\Nat \to \Nat}
\qquad\text{and}\qquad
\isterm{f : \Nat \to \Nat,\ \x:\Nat}{x}{\Nat}.
\end{equation*}
In a system with strengthening, we could immediately see that $x$ is unnecessary in the
first judgment and $f$ in the second. The loss of strengthening is inconvenient enough
that we restore it by explicitly keeping track of dependencies on the assumptions in the
context.

In the implementation we use \emph{terms with assumptions}, which are ordinary terms that
have every subterm annotated with a set of variables, called the \emph{assumptions},
indicating explicitly which part of the context a subterm depends on. Thus
$\isterm{\G}{\ttt^\alpha}{\AA^\beta}$ means that we may restrict $\G$ to variables
in~$\alpha$ to obtain a smaller context $\ctxrestrict{\G}{\alpha}$ in which it is still
possible to show that $\ttt$ has type $\AA$. Similarly, $\ctxrestrict{\G}{\beta}$ suffices
to derive the judgment that~$\AA$ is a type. The types and terms appearing in the context
are themselves annotated with assumptions, which endows contexts with the structure of
directed acyclic graphs. (In the implementation they are stored as such.)

This means that Andromeda can compose and decompose judgments without information loss.
The application above will be recorded internally as:
\begin{equation*}
\isterm
  {f:(\Nat^\emptyset \to {\Nat^\emptyset})^\emptyset,\, \x:\Nat^\emptyset}
  {(\app{f^{\{f\}}}{\_}{\Nat^\emptyset}{\Nat^\emptyset}{x^{\{x\}}})^{\{f,x\}}}
  {\Nat^\emptyset}.
\end{equation*}
and it is straightforward to recover the two original sub-judgments.

Constants from the signature are not included in assumption sets, since they are
omnipresent anyhow.

\paragraph*{Context joins}

The standard rules of inference require the contexts of the premises to match, for
instance the application rule $\rulename{term-app}$ does not allow a change of the
context:
\begin{equation*}
  \infer
  {\isterm{\G}{\sss}{\Prod{x}{\AA} \BB} \\
   \isterm{\G}{\ttt}{\AA}
  }
  {\isterm{\G}{\app{\sss}{\x}{\AA}{\BB}{\ttt}}{\subst{\BB}{\x}{\ttt}}}
\end{equation*}
If we implemented the rule exactly as is, the user would have to plan dependence on
hypotheses carefully in advance, which is impractical. Instead, we rely on admissibility
of weakening to enlarge contexts as necessary. In every inference rule we accept premises
with arbitrary contexts that are then \emph{joined} to form a single extended context,
for instance the application rule becomes
\begin{equation*}
  \infer
  {\isterm{\G}{\sss}{\Prod{x}{\AA} \BB} \\
   \isterm{\D}{\ttt}{\AA}
  }
  {\isterm{\ctxjoin{\G}{\D}}{\app{\sss}{\x}{\AA}{\BB}{\ttt}}{\subst{\BB}{\x}{\ttt}}}
\end{equation*}
The context join $\ctxjoin{\G}{\D}$ is the smallest context that extends both $\G$ and
$\D$. In terms of directed acyclic graphs it is just the union of graphs. A context join
fails if there is a hypothesis that has different types in $\G$ and $\D$, or if the join
would create a cyclic dependency of hypotheses. In practice such failures are infrequent;
\lstinline{λ}, \lstinline{Π}, and \lstinline{assume} create globally fresh object-level variables,
so there is no direct way to create two contexts with the same variable at different types.\footnote{An indirect method to obtain unjoinable contexts is to take a single judgment with the context $X{:}\Type, x{:}X$ and explicitly substitute in two different ways, replacing $X$ with two distinct types.}
\medskip

Andromeda automatically tracks assumption sets and contexts. Even though the implementation
makes an effort to keep them small, they may not be unique or minimal: they merely reflect
a history of how judgments were constructed.

%%% Local Variables:
%%% mode: latex
%%% TeX-master: "main"
%%% End:

\section{The Andromeda meta-language}
\label{sec:meta-language}

The Andromeda meta-language (AML) is a programming language in the style of ML~\cite{standard-ml}.
We review its structure and capabilities, focusing on the parts that are peculiar to
Andromeda. For constructs that are standard in the ML-family of languages, such as type
definitions, \lstinline|let|-bindings, recursive functions, etc., we refer the reader to
the Andromeda reference page.\footnote{\url{http://www.andromeda-prover.org/meta-language.html}}

In order to distinguish the expressions of AML from the expressions of the object-level type
theory, we refer to the former as \emph{computations} to emphasize that their evaluation
may have side effects (such as printing things on the screen), and to the latter as
\emph{(type-theoretic) terms}. We refer to the types of AML as \emph{ML-types}.

Keep in mind that the ML-level computations can never enter the object-level terms, as the
nucleus knows nothing about~AML. What looks like AML code inside an object-level term is
\emph{always} just AML code that constructs a judgment. For example, a pattern match
inside a $\lambda$-abstraction, \lstinline|λ(x:A), match … end|, is a computation that evaluates the
\lstinline|match| statement immediately to obtain an object-level term, which is then
abstracted. In contrast, the ML-level function \lstinline|fun x ⇒ match … end| does suspend
the evaluation of its body.

\subsection{ML-types}
\label{sec:ml-types}

AML is equipped with static type inference in the style of Hindley-Milner
parametric polymorphism~\cite{hindley-milner}. It supports definitions of
parametric ML-types, including inductive types. The only non-standard aspect of
the ML-type inference arises from the fact that application is overloaded, as it
is used both for invoking ML-level application and for building object-level
applications. For instance the type of \lstinline|f|, defined by
\begin{lstlisting}
let f x y = x y
\end{lstlisting}
could be either $\judgTy \to \judgTy \to \judgTy$ or
$(\alpha \to \beta) \to \alpha \to \beta$; in such cases the
inferred type constraints
are postponed until we are sure that \lstinline{x} will be a judgment or
that at least one of \lstinline{x} and \lstinline{y} will not. This strategy
works well in practice, with only the occasional application constraint remaining
unresolved at the top level.

\subsection{Pattern matching}
\label{sec:pattern-matching}

AML pattern matching in \lstinline|match| statements and \lstinline|let|-bindings is more flexible than that
of Standard~ML and related languages. AML patterns need not be linear (i.e.,
a pattern variable may appear several times in a pattern) and variables may be
interpolated into patterns. Pattern variables are prefixed with \lstinline|?| so
that they can be  distinguished from
interpolated variables.
For example,
\begin{itemize}
\item the pattern \lstinline|(?x, ?y)| matches any pair,
\item the pattern \lstinline|(?x, ?x)| matches a pair whose components are equal,
\item the pattern \lstinline|(?x, y)| matches a pair whose second component equals the
  value of~\lstinline|y|.
\end{itemize}
Equality in pattern matching always means syntactic identity ($\alpha$-equivalence in the
case of object-level terms), not arbitrary judgmental equalities. The flexibility of
pattern matching is handy when we match on values of type $\judgTy$; see~\S\ref{sec:datatype-judgments},
where we also discuss patterns for deconstruction of typing judgments.

\subsection{Operations and handlers}
\label{sec:operations-handlers}

During evaluation of a computation of ML-type $\judgTy$ the interpreter may need evidence of
equality between two types (in order to present it to the nucleus), which it gets by passing control back to user code, together
with information on what needs to be done, and how to resume the evaluation once the
evidence is obtained. To accomplish this, AML is equipped with algebraic operations and
handlers~\cite{handlers} in the style of Eff~\cite{eff-lang}. We
recommend~\cite{pretnar-tutorial,eff-lang} for background
reading, and give just a quick overview here. A more detailed discussion on the use of
algebraic operations and handlers for the purposes of computing judgments can be found in
\S\ref{sec:datatype-judgments}.

One way to think of an operation is as a generalized resumable exception: when an
operation is invoked it ``propagates'' outward to the innermost handler that handles it.
The handler may then perform an arbitrary computation, and using $\cyield{c}$ it
may resume the execution at the point at which the operation was invoked, yielding
the value of~$c$ as the result of the operation. Similarly, we can think of a handler as a generalized exception
handler, except that it handles one or more operations, as well as values (computations
which do not invoke an operation). An example of handlers in action is given in~\S\ref{sec:using-equality-hints}.

\subsection{The datatype $\judgTy$}
\label{sec:datatype-judgments}

In Andromeda the user \emph{always works with an entire judgment
  $\isterm{\G}{\ttt}{\AA}$,} and never a bare term $\ttt$. Similarly a type $\AA$ never
stands by itself, but always in a judgment $\isterm{\G}{\AA}{\Type}$. The judgments are
represented by values of a special primitive type $\judgTy$.

\paragraph*{Judgment forms}

The OCaml interface for the nucleus uses distinct abstract datatypes to represent the
different judgment forms. These distinctions are not visible to the user, because AML exposes all
forms through the single datatype $\judgTy$ whose values are judgments of the form
$\isterm{\G}{\ttt}{\AA}$. This is possible because $\Type : \Type$ and equality reflection
let us express the three forms
$\istype{\G}{\AA}$,
$\eqterm{\G}{\sss}{\ttt}{\AA}$, and
$\eqtype{\G}{\AA}{\BB}$
as
$\isterm{\G}{\AA}{\Type}$,
$\isterm{\G}{p}{\JuEqual{\AA}{\sss}{\ttt}}$, and
$\isterm{\G}{q}{\JuEqual{\Type}{\AA}{\BB}}$, respectively.

We hope the user finds it simpler to access all object-level entities in
a uniform way. On the other hand, having more precise judgment types in AML
would help catch potential errors. We discuss this particular design choice in
\S\ref{sec:remove-type-in-type}.

In AML no direct datatype constructors for~$\judgTy$ are available. (Even at
the level of OCaml implementation the datatype constructors are invisible outside the
nucleus.) Instead, the user may invoke primitive computations of type $\judgTy$ that
\emph{look like} term constructors, but really correspond to inference rules of type
theory. For instance, an application $\capp{c_1}{c_2}$, where $c_1$ and $c_2$ are
computations of type $\judgTy$, computes an instance of the $\rulename{term-app}$ rule
(actually, the version with context joins). The user does not have to provide the explicit
typing annotations on the application, as these are derived using a bidirectional typing
strategy, as described next.

\paragraph*{Inferring and checking modes of evaluation}

There are two
modes of AML evaluation, \emph{inferring} and \emph{checking}. In inferring mode the type of
the result is unconstrained. In checking mode the type is prescribed in advance: there is
given a type $\AA$ (or more precisely, a judgment $\isterm{\G}{\AA}{\Type}$) and the
computation must evaluate to a judgment of the form $\isterm{\D}{\ttt}{\AA}$ where $\D$
extends $\G$.

For instance, an application $\capp{c_1}{c_2}$ is evaluated in inferring mode as follows.
First $c_1$ is evaluated in inferring mode to $\isterm{\G}{\sss}{\Prod{\x}{\AA}{\BB}}$ (we
discuss what happens if the type of $\sss$ is not a product below), then $c_2$ is
evaluated in checking mode at type~$\AA$ to $\isterm{\D}{\ttt}{\AA}$, and the result is
$\isterm{\ctxjoin{\G}{\D}}{\app{\sss}{\x}{\AA}{\BB}{\ttt}}{\subst{\BB}{\x}{\ttt}}$.

\paragraph*{Judgment computations}

The following primitives for computing judgments are provided:
\begin{itemize}
\item Primitives for term and type formation:
  \begin{gather*}
    \cType \qquad
    \cProd{x}{c_1} c_2 \qquad
    \capp{c_1}{c_2} \qquad
    \clam{x}{c_1} c_2 \qquad
    \cEq{c_1}{c_2} \qquad
    \crefl{c}.
  \end{gather*}
  Note that the notation $\cEq{c_1}{c_2}$ is used for the equality type, rather than for
  judgmental equality, which the user never writes down explicitly. We emphasize again
  that these are \emph{not} datatype constructors for forming terms and types of the
  object-level type theory, but primitive \emph{computations}, with possible side effects,
  that build \emph{judgments} from sub-judgments by passing through the nucleus.

\item Type ascription $\cascribe{c_1}{c_2}$, which first evaluates $c_2$ to
  $\isterm{\G}{\AA}{\Type}$ and then evaluates $c_1$ in checking mode at type $\AA$.

\item Top-level \lstinline|constant| declarations, which introduce new constants.

\item
  The computation $\cassume{x}{c_1} c_2$, which
  evaluates~$c_1$ in inferring mode to~$\isterm{\G}{\AA}{\Type}$, and then $c_2$ with $x$
  \lstinline|let|-bound to $\isterm{\ctxextend{\G}{x_i}{\AA}}{x_i}{\AA}$, where $x_i$ is a
  freshly generated name. This should not be confused with constant declarations: a
  constant is an omnipresent part of the signature, while an assumption is local to a
  judgment in which it appears, and is tracked in assumption sets. Furthermore, we
  may replace an assumption with a term, using the substitution primitive, but not a
  constant.

\item Substitution $\csubst{c_1}{x}{c_2}$, which replaces $x$ with the value of $c_2$ in
  the value of $c_1$, assuming the types match.

\item The computation $\coccurs{c_1}{c_2}$, which evaluates $c_1$ to a judgment
  $\isterm{\D}{\x}{\AA}$ and $c_2$ to a judgment $\isterm{\G}{\ttt}{\BB}$, and checks
  whether $x$ appears in $\G$. It returns $\mathtt{None}$ if not, and
  $\mathtt{Some}(\isterm{\E}{\CC}{\Type})$ if $\x$ appears in $\Gamma$ as a variable of
  type $\CC$.

\item Computations that generate witnesses for the $\beta$-rule and the congruence rules.
  There are no primitive computations for extensionality rules $\rulename{eq-eta}$ and
  $\rulename{prod-eta}$ because they can be declared with a constant by the user. Indeed,
  we do so in the standard library.

\item The computation $\cnatural{c}$, which witnesses uniqueness of typing. It evaluates $c$
  to a judgment $\isterm{\G}{\ttt}{\AA}$ and outputs a witness for equality of $\AA$ and
  the natural type $\naturalTy{\ttt}$ of~$\ttt$. The witnesses are needed when a tactic
  deconstructs a term and puts it back together, thus obtaining the original term at its
  natural type.
\end{itemize}

\paragraph*{Judgment patterns}

Apart from computations that form judgments, we also need flexible ways of analyzing and
deconstructing them. In AML this is done with the $\mathtt{match}$ statement and judgment
patterns of the form $\jpatt{p_1}{p_2}$, where $p_2$ may be omitted, and $p_1$ and $p_2$
are among the following:
\begin{itemize}
\item Anonymous pattern \lstinline|_|, pattern variables $\pvar{x}$, and interpolated
  variables $x$.
\item Patterns for matching terms and types:
  \begin{gather*}
    \cType \qquad
    \cProd{\pvar{x}}{p_1} p_2 \qquad
    \capp{p_1}{p_2} \qquad
    \clam{\pvar{x}}{p_1} p_2 \qquad
    \cEq{p_1}{p_2} \qquad
    \crefl{p}.
  \end{gather*}
  Note that the patterns for products and abstractions ``open up'' the binders so that it
  is possible to pattern-match under the binders; the judgment matched
  by $p_2$ can have the bound variable in its context.
\item Patterns for matching free variables $\patom{\pvar{x}}$ and constants
  $\pconst{\pvar{x}}$.
\end{itemize}
More precisely, when $\isterm{\G}{\ttt}{\AA}$ is matched with $\jpatt{p_1}{p_2}$, the
term~$\ttt$ is matched with~$p_1$ and the type~$\AA$ with~$p_2$. Assuming the match
succeeds, the pattern variables in~$p_1$ and~$p_2$ are bound to sub-judgments that are
obtained through inversion lemmas~\S\ref{sec:implementation-of-type-theory}.
The contexts of the sub-judgments are kept minimal thanks to assumption sets. Examples of
pattern matching are shown in Appendix~\ref{sec:auto-tactic}.

Pattern matching is always executed at the AML level; patterns and the
\lstinline{match} statements exist only as computations, and are not part of the
object-level terms. To highlight this point, we show the difference between a
\lstinline{match} inside $\lambda$-abstraction and an AML function. Assuming a type
\lstinline{A} with two constants \lstinline{a, b : A} and an endofunction \lstinline{f : A → A},
the computation
\begin{lstlisting}
(λ (x : A), match x with
              | ⊢ ?g ?y ⇒ y
              | ⊢ _     ⇒ b
            end             ) (f a)
\end{lstlisting}
evaluates to the judgment \lstinline{⊢ (λ (x : A), b) (f a) : A}, while
\begin{lstlisting}
(fun x ⇒ match x with
           | ⊢ ?g ?y ⇒ y
           | ⊢ _     ⇒ b
         end             ) (f a)
\end{lstlisting}
evaluates to the judgment \lstinline{⊢ a : A}. In the former case matching occurred inside
the abstraction, so \lstinline{x} evaluated to \lstinline{⊢ x : A} and the second clause
matched; in the latter case matching took place when the function was applied, so
\lstinline{x} was bound to \lstinline{⊢ f a : A} and the first clause matched.

The AML interpreter matches a judgment against a pattern by first asking the nucleus to
invert the judgment. The nucleus returns information about which inversion was used and
what constituent parts it produced, from which the interpreter calculates whether the
pattern matches and how. If there are sub-patterns, the process continues recursively.
Pattern matching uses syntactic equality (up to $\alpha$-equivalence) and never triggers
any operations, although an inexhaustive \lstinline{match} may fail.

At present there are no judgment patterns for analyzing the context of a judgment.
Instead, the primitive computation $\ccontext{c}$ evaluates $c$ to a judgment
$\isterm{\G}{\ttt}{\AA}$ and gives the list of all hypotheses in $\G$, sorted so that each
hypothesis is preceded by its dependencies.

\paragraph*{Equality checks and coercions}

AML only verifies syntactic equality automatically. It delegates any other equality
$\eqterm{\G}{\sss}{\ttt}{\AA}$ by triggering the operation
$\opEqual{(\isterm{\G}{\sss}{\AA})}{(\isterm{\G}{\ttt}{\BB})}$, which passes control back
to the user-level AML code. The operation may go unhandled, in which case an error is
reported, or it may be intercepted by a handler in the user code. The handler may do whatever it wants,
but the intended use is for it to attempt to calculate evidence of the given equality. The
handler yields $\mathtt{None}$ if it fails to compute the evidence (in which case the
interpreter reports an error), or
$\mathtt{Some}(\isterm{\D}{\xi}{\JuEqual{\AA}{\sss}{\ttt}})$ if it finds a witness
$\isterm{\D}{\xi}{\JuEqual{\AA}{\sss}{\ttt}}$. Note that the handler is itself a piece of
AML code that may recursively trigger further operations and handling thereof.

Apart from equality checking, there are other situations in which the AML interpreter
triggers an operation:
\begin{itemize}
\item It may happen that AML needs to know why a given type $\isterm{\G}{\AA}{\Type}$ is
  equal to a product type. Unless $\AA$ is already syntactically equal to a product type,
  the interpreter triggers an operation $\opAsProd{(\isterm{\G}{\AA}{\Type})}$. It expects a
  handler to yield $\mathtt{None}$ upon failure, or
  $\mathtt{Some}(\isterm{\D}{\xi}{\JuEqual{\Type}{\AA}{\Prod{\x}{\BB}{\CC}}})$ witnessing
  that $\AA$ is equal to a product type.

\item Similarly, if AML needs to know why $\isterm{\G}{\AA}{\Type}$ is equal to an
  equality type, it triggers an operation $\opAsEq{(\isterm{\G}{\AA}{\Type})}$. It expects
  the handler to yield $\mathtt{None}$ or
  $\mathtt{Some}(\isterm{\D}{\xi}{\JuEqual{\Type}{\AA}{\JuEqual{\BB}{\sss}{\ttt}}})$.

\item If an inferring term evaluates to $\isterm{\G}{\ttt}{\AA}$ in checking mode at type
  $\isterm{\D}{\BB}{\Type}$, the interpreter does \emph{not} ask for evidence that $\AA$
  and $\BB$ are equal, but instead triggers the operation
  $\opCoerce{(\isterm{\G}{\ttt}{\AA})}{(\isterm{\D}{\BB}{\Type})}$ that gives the user
  code an opportunity to replace $\ttt$ with a value of type~$\BB$. The handler must
  yield:
  \begin{itemize}
  \item $\mathtt{NotCoercible}$ to indicate failure to coerce $\ttt$ to $\BB$,
  \item $\mathtt{Convertible} (\isterm{\E}{\xi}{\JuEqual{\Type}{\AA}{\BB}})$ to indicate
    that $\AA$ and $\BB$ are equal, so that AML may apply conversion to $\ttt$, or
  \item $\mathtt{Coercible}(\isterm{\E}{\sss}{\BB})$ to have $\ttt$ replaced with $\sss$.
  \end{itemize}
  This mechanism allows the user to implement various strategies for coercion of values,
  and control them completely through handlers.

\item If the head $c_1$ of an application $\capp{c_1}{c_2}$ evaluates to a term
  $\isterm{\G}{\ttt}{\AA}$ where $\AA$ is not a product type, the interpreter asks the user code to
  convert $\ttt$ to a function by triggering the operation
  $\opCoerceFun{(\isterm{\G}{\ttt}{\AA})}$. The handler should yield
  $\mathtt{NotCoercible}$,
  $\mathtt{Convertible}(\isterm{\D}{\xi}{\JuEqual{\Type}{\AA}{\Prod{\x}{\BB} \CC}})$, or
  $\mathtt{Coercible}{(\isterm{\D}{\sss}{\Prod{\x}{\BB} \CC})}$, as the case may be.
\end{itemize}

\subsection{References and dynamic variables}
\label{sec:refs-and-dyno}

As a convenience, AML provides ML-style mutable references. They are used to store the
current state of implicit arguments in the standard library
(see~\S\ref{sec:implicit-arguments}).

AML also supports \emph{dynamic variables}. These are globally defined mutable values with
dynamic binding discipline. A dynamic variable $x$ is declared and initialized with the
top-level command $\cdynamic{x}{c}$. The computation $\cnow{x}{c_1}{c_2}$ changes $x$ to
the value of $c_1$ locally in the computation of $c_2$.

AML maintains a dynamic variable \lstinline{hypotheses}. It is a list of judgments that
plays a role in evaluation of computations under binders. To evaluate $\clam{\x}{\AA} c$,
AML generates a fresh variable $\x_i$ of type $\AA$, binds $\x$ to the judgment
$\isterm{\x_i:\AA}{\x_i}{\AA}$, prepends it to \lstinline{hypotheses}, evaluates~$c$, and
abstracts $\x_i$ to get the final result. By accessing \lstinline{hypotheses} the
computation~$c$ may discover under what binders it is evaluated. For example,
in~\S\ref{sec:system-design} the handler for the \lstinline{auto} tactic searched
\lstinline{hypotheses} for ways of inhabiting a type.

The standard library uses dynamic variables \lstinline{betas}, \lstinline{etas},
\lstinline{hints}, and \lstinline{reducing} to store $\beta$-hints, $\eta$-hints, general
hints, and reduction directives. It is important for these variables to
follow a dynamic binding discipline so that \emph{local} equality hints work correctly
(see~\S\ref{sec:equality-hints}).

%%% Local Variables:
%%% mode: latex
%%% TeX-master: "main"
%%% End:

\section{Soundness of Andromeda}
\label{sec:soundness-andromeda}

Soundness in Andromeda has both theoretical and engineering aspects.

Theoretical soundness pertains to the differences between the original type theory,
(Appendix~\ref{sec:rules-type-theory}) and the type theory implemented in the nucleus
(\S\ref{sec:implementation-of-type-theory}), which uses assumption sets, context joins,
and natural types. In the following we write $\sss^\sigma$ for a term $\sss$ decorated
with assumptions $\sigma$, i.e., if we remove assumptions sets from the decorated term
$\sss^\sigma$ we get the ordinary term~$\sss$. We follow a similar convention for types
and context.

\begin{claim}
  \label{claim:type-theory}%
  Given a context $\Gamma$, a term $\sss$ and a type $\AA$:
  \begin{enumerate}
  \item If $\isterm{\G}{\sss}{\AA}$ is derivable in the original type theory, then
    $\istermI{\D^\delta}{\sss^\sigma}{\AA^\alpha}$ is derivable for some $\Delta^\delta$,
    $\sss^\sigma$, and $\AA^\alpha$ such that $\D$ is a subcontext of $\G$.
  \item If $\istermI{\G^\gamma}{\sss^\sigma}{\AA^\alpha}$ is derivable in the implemented
    type theory, then $\isterm{\G}{\sss}{\AA}$ is derivable in the original type theory.
  \end{enumerate}
\end{claim}

\noindent
We cannot call the statement a theorem because we have not yet proved it in detail.
We leave the task as future work for this progress report, and note that we do not anticipate a particularly
enlightening or difficult proof, just the usual grinding of cases by structural induction.
The most interesting part of the proof will likely by the formalization of the implemented
type theory from \S\ref{sec:implementation-of-type-theory}, which we have postponed
because it has been regularly modified as we gained experience with the implementation.

The second aspect of soundness is an engineering question: how do we know that the
implementation of Andromeda works as intended?

\begin{claim}
  \label{claim:implementation}%
  If Andromeda evaluates a computation to a judgment, then the judgment
  is derivable from the implemented type theory
  with respect to the signature containing all the
  constants declared by the user.
\end{claim}

Let us reiterate the design choices we have made to give credence to the claim. In the
OCaml implementation the datatypes representing the judgment forms are all abstract and
kept opaque by an interface to a small trusted OCaml module, the \emph{nucleus}. We rely
on the soundness of OCaml's type system to ensure that the untrusted remainder of the
system cannot forge new values of these abstract types.\footnote{If we were to reimplement
  the system in an unsafe language such as C, or if we lacked faith in OCaml, additional
  mechanisms such as cryptographic signatures could be used.} The nucleus is kept as
simple as possible, and it only supports very straightforward type-theoretic constructions
which directly correspond to applications of inference rules and admissible rules.
Everything else, the AML type inference, the core AML intepreter, the implementation of operations and
handlers, and the user code, is on the other side of the barrier and does not influence
soundness. In particular, the nucleus does not know anything about AML at all, does not
trigger operations, and no AML code can ever enter the object-level terms (so there
is no question about having pattern matching, exotic terms involving AML code, or any
other part of AML at the object level).

Formally verifying the 1900 lines of the nucleus code is still a tall order to handle, and
at present we have no plans to do it. Once we have formulated the implemented type theory,
a careful code review of the nucleus will probably unearth some bugs, and hopefully not
very many!

There is a third kind of soundness, namely the consistency of the underlying type theory.
Obviously, since we included $\Type : \Type$ the theory is at present inconsistent in the
sense that all types are inhabited and all judgmental equalities derivable. As soon as we
remove $\Type : \Type$ the theory becomes consistent, since what remains are just bare
products and equality types with reflection, and these are consistent in virtue of having
a model (such as the hereditarily finite sets). We discuss removal of
$\Type : \Type$ in \S\ref{sec:remove-type-in-type}.

%%% Local Variables:
%%% mode: latex
%%% TeX-master: "main"
%%% End:

\section{The standard library}
\label{sec:stand-libr}

To test the viability of our design we implemented a small standard library in AML. By
design, anything that is implemented in AML is safe: it may not work as expected, or
diverge, but it will never produce an invalid judgment, or derive an invalid equality. (Of
course, this does not say much until we have dealt with $\Type : \Type$.)

\subsection{Equality checking}
\label{sec:equality-checking}

The most substantial part of the library is a user-extensible equality checking algorithm
with rudimentary support for implicit arguments, based on similar
ones by Stone and Harper~\cite{harper:stone} and Coquand~\cite{coquand:lf91}. It
computes a witness of equality $\eqterm{\G}{\sss}{\ttt}{\AA}$ in two phases:
\begin{itemize}
\item The \emph{type directed} phase computes the weak head-normal form (whnf) of type~$A$ to see
  whether any extensionality rules apply. For instance, if $\AA$ normalizes to a product
  $\Prod{\x}{\BB}{\CC}$, the algorithm applies function extensionality
  $\rulename{prod-eta}$ to reduce the equality to
  $
   \eqterm{\ctxextend{\G}{\y}{\BB}}
          {\app{\sss}{\x}{\BB}{\CC}{\y}}
          {\app{\sss}{\x}{\BB}{\CC}{\y}}{\subst{\BB}{\x}{\y}}
  $
  at a smaller type. Similarly, if $\AA$ is an equality type the equality checks succeeds
  immediately by uniqueness of equality proofs $\rulename{eq-eta}$.
  Extensionality rules including $\rulename{prod-eta}$ and $\rulename{eq-eta}$ are user defined
  (see~\S\ref{sec:equality-hints}).
\item Once the type-directed phase simplifies the type so that no further extensionality
  rules apply, the \emph{normalization phase} computes the weak head-normal forms of
  $\sss$ and $\ttt$ and compares them structurally, which generates new equality problems
  involving subterms.
\end{itemize}

The equality checking algorithm relies on the computation of weak head-normal forms of
terms, which is also implemented by the standard library. Given a term
$\isterm{\G}{\ttt}{\AA}$, the library computes a witness
$\isterm{\G}{\xi}{\JuEqual{\AA}{\ttt}{\ttt'}}$ where $\ttt'$ is in weak head-normal form.
It does so by chaining together a sequence of \emph{computation rules} using transitivity
of equality. By default the only computation rule is $\rulename{prod-beta}$ for reducing
$\beta$-redices, but the user may install additional rules as explained in~\S\ref{sec:equality-hints}.

\subsection{Equality hints}
\label{sec:equality-hints}

The equality checking algorithm can be extended by the user with new rules, which we call
\emph{equality hints}. There are three kinds:
\begin{itemize}
\item an \emph{$\eta$-hint}, or an extensionality hint, is a term whose type has the form
  \begin{equation*}
    \Prod{\x_1}{\AA_1} \cdots \Prod{x_n}{\AA_m}
    \Prod{\y_1}{\BB} \Prod{y_2}{\BB}
    \JuEqual{\CC_1}{\ttt_1}{\sss_1} \to \cdots \to
    \JuEqual{\CC_m}{\ttt_m}{\sss_m} \to
    \JuEqual{\BB}{y_1}{y_2}.
  \end{equation*}
  It is a universally quantified equation with equational preconditions, where the
  left-hand and the right-hand side of the equation are distinct variables. The equality
  checking algorithm matches such a hint against the goal. If the match succeeds, the goal
  is reduced to deriving the preconditions.
\item a \emph{$\beta$-hint}, or a computation hint, is a term whose type is a universally
  quantified equation
  \begin{equation*}
    \Prod{\x_1}{\AA_1} \cdots \Prod{x_n}{\AA_m}
    \JuEqual{\CC}{\sss}{\ttt}.
  \end{equation*}
  The weak head-normal form algorithm matches the left-hand side $s$ of the equation
  against the term. If the match succeeds, it performs a reduction step from $s$ to $t$.
\item a \emph{general hint} is a term whose type is a universally quantified equation
  \begin{equation*}
    \Prod{\x_1}{\AA_1} \cdots \Prod{x_n}{\AA_m}
    \JuEqual{\CC}{\sss}{\ttt}.
  \end{equation*}
  The equality checking algorithm matches such a hint against the goal during the type
  directed phase to see whether it can immediately dispose of the goal.
\end{itemize}
In addition, the user may give a \emph{reduction strategy} for a given constant by
specifying which of its arguments should be reduced eagerly. This is necessary for the
equality checking algorithm to work correctly when we introduce new eliminators. For
instance, when we axiomatize simple products $\AA \times \BB$, the extensionality rule
\begin{multline*}
  \Prod{\AA}{\Type}
  \Prod{\BB}{\Type}
  \Prod{x, y}{\AA \times \BB} \\
  \JuEqual{\AA}{\mathtt{fst}\,\AA\,\BB\,x}{\mathtt{fst}\,\AA\,\BB\,y} \to
  \JuEqual{\AA}{\mathtt{snd}\,\AA\,\BB\,x}{\mathtt{snd}\,\AA\,\BB\,y} \to
  \JuEqual{\AA \times \BB}{x}{y}
\end{multline*}
only works correctly if we also specify that the normal form of a projection
$\mathtt{fst}\,\AA\,\BB\,\ttt$ should have $\ttt$ normalized, and similarly for
$\mathtt{snd}$. Another example is the recursor for natural numbers, which should eagerly
reduce the number at which it is applied.

Examples of equality hints and uses of reduction strategies will be shown in
\S\ref{sec:examples}. Let us only remark that the hints and reduction strategies may be
installed locally, even under a binder using a temporary equality assumption, and that the
user is free to install whatever hints they wish, including ones that break
completeness of the algorithm.
However, as long as hints are confluent and strongly normalizing, the algorithm
behaves sensibly.

\subsection{Implicit arguments}
\label{sec:implicit-arguments}

The standard library provides basic support for implicit arguments. In other
systems these are usually implemented with meta-variables, which are not
available in AML. In their place, we use ordinary fresh variables generated
using the \lstinline{assume} construct. We refer to these as \emph{implicit
variables}. We collect constraints through the operations and handlers mechanism,
and resolve them using a simple first-order unification procedure.

More precisely, in the standard library we declare operations
\begin{lstlisting}
operation ? : judgment
operation resolve : judgment → judgment
\end{lstlisting}
The user may place \verb|?| anywhere where they want the term to be derived
automatically, and call $\mathtt{resolve}\,c$ to replace the implicit variables
with their derived values in the judgment computed by~$c$.

The handler provided by the library keeps a list of implicit variables it has introduced
so far, as well as their types and known solutions. The operation $\verb|?|$ may be
triggered either in checking or inferring mode. In checking mode at type~$\AA$ and under
binders $\x_1 : \BB_1, \ldots, \x_n : \BB_n$, the handler introduces a fresh implicit
variable $M$ of type $\Prod{\x_1}{\BB_1} \ldots \Prod{\x_n}{\BB_n} \AA$ and yields
$M\,x_1 \ldots x_n$. In inferring mode the type $\AA$ is not available. For simplicity, at
present the library will report an error, although it might be better to create an implicit variable
for the $\AA : \Type$.

During equality checking we may discover that $M\,x_1 \ldots x_n$ should be equal to a
term~$t$ in which $M$ does not occur. In this case, using \lstinline|assume| again, the
handler generates a term $\xi : \JuEqual{}{M}{\lambda x_1 \ldots x_m \,.\, t}$, stores it,
and also installs it as a $\beta$-hint, so that subsequent equality checks take it into
account.

The operation $\mathtt{resolve}\;c$ is used to replace the implicit variables with their
inferred values in the judgment computed by~$c$. Such replacement does not happen
automatically because the library cannot guess when is the best moment for doing so.
It may be necessary to evaluate several computations before all the implicit
variables become known, so we let the user control when resolution should happen.

While we feel quite encouraged by our implementation of equality checking, the
implicit arguments feel a bit heavy-handed, and are quite slow. They are a satisfactory
proof of concept and a demonstration of the flexibility of operations and
handlers, but we need to improve it quite a bit before it becomes useful.

%%% Local Variables:
%%% mode: latex
%%% TeX-master: "main"
%%% End:

\section{Examples}
\label{sec:examples}

In this section we show Andromeda at work through several examples.

\subsection{Proving equality with handlers}
\label{sec:using-equality-hints}

As explained in~\S\ref{sec:operations-handlers}, when AML is faced with proving a
non-trivial equality, it delegates it to user code by triggering the operation
\lstinline|equal|. To see how this works, let us walk through a computation that
constructs a term witnessing symmetry of equality (without the standard library
installed):
\begin{lstlisting}
λ (A : Type) (x y : A) (p : x ≡ y),
  (handle
    refl x : y ≡ x
   with
   | equal x y ⇒ yield (Some p)
   end)
\end{lstlisting}
The $\lambda$-abstraction introduces a type \lstinline{A}, elements \lstinline{x},
\lstinline{y} of type \lstinline{A}, and a witness \lstinline{p} of equality between
\lstinline{x} and \lstinline{y}. Next, the type ascription is evaluated, with the
enveloping handler installed. First \lstinline{y ≡ x} is evaluated to the equality type
$\JuEqual{\mathtt{A}}{\mathtt{y}}{\mathtt{x}}$ and then \lstinline{refl x} is evaluated in
checking mode at this type. This triggers a sub-computation of \lstinline{x} and verification that
\lstinline{x} equals \lstinline{y} (and a trivial equality check that \lstinline{x} equals to itself).
At this point AML triggers the operation \lstinline{equal x y}, asking for evidence of equality.
The enveloping handler intercepts the operation and yields the evidence \lstinline{p}.

The result of the computation is displayed by Andromeda without typing annotations and
assumption sets as
\begin{lstlisting}
⊢ λ (A : Type) (x : A) (y : A) (_ : x ≡ y), refl x
  : Π (A : Type) (x : A) (y : A), x ≡ y → y ≡ x
\end{lstlisting}

\subsection{Dependent sums}
\label{sec:dependent-sums}

Our second example shows how to axiomatize dependent sums. This time we use the standard
library and rely on its equality checking. We start by postulating the type and term
constructors:
\begin{lstlisting}
constant Σ : Π (A : Type) (B : A → Type), Type
constant existT : Π (A : Type) (B : A → Type) (a : A), B a → Σ A B
\end{lstlisting}
Next, we postulate the projections and tell the standard library that that the third
argument of a projection should be evaluated eagerly, so that we get a working
extensionality rule later on:
\begin{lstlisting}
constant π₁ : Π (A : Type) (B : A → Type), Σ A B → A
now reducing = add_reducing π₁ [lazy, lazy, eager]

constant π₂ : Π (A : Type) (B : A → Type) (p : Σ A B), B (π₁ A B p)
now reducing = add_reducing π₂ [lazy, lazy, eager]
\end{lstlisting}
It remains to postulate equalities, and install them as hints. The $\beta$-rules are
straightforward, except that we must install the $\beta$-rule for the first projection
before we postulate the second projection, or else Andromeda does not know why the second
projection is well typed:
\begin{lstlisting}
constant π₁_β :
  Π (A : Type) (B : A → Type) (a : A) (b : B a),
    (π₁ A B (existT A B a b) ≡ a)

now betas = add_beta π₁_β

constant π₂_β :
  Π (A : Type) (B : A → Type) (a : A) (b : B a),
    (π₂ A B (existT A B a b) ≡ b)

now betas = add_beta π₂_β
\end{lstlisting}
Similarly, to convince Andromeda that the extensionality rule is well typed, we need to
install a \emph{local} hint, as follows (the function \lstinline{symmetry} is part of the
standard library and it computes the symmetric version of an equality):
\begin{lstlisting}
constant Σ_η :
  Π (A : Type) (B : A → Type) (p q : Σ A B)
    (ξ : π₁ A B p ≡ π₁ A B q),
    now hints = add_hint (symmetry ξ) in
      π₂ A B p ≡ π₂ A B q → p ≡ q

now etas = add_eta Σ_η
\end{lstlisting}

\subsection{Natural numbers}
\label{sec:natur-numb}

The standard library provides functions for calculating weak head-normal forms that can be used
as a computation device at the level of type-theoretic terms. We show how this is done
by axiomatizing natural numbers and computing with them.

We postulate the type of natural numbers and its constructors
\begin{lstlisting}
constant nat : Type
constant O : nat
constant S : nat → nat
\end{lstlisting}
and the induction principle
\begin{lstlisting}
constant nat_rect : ∏ (P : nat → Type),
  P O → (∏ (n : nat), P n → P (S n)) → ∏ (m : nat), P m
\end{lstlisting}
The weak head-normal form of the eliminator should have the fourth argument
normalized:
\begin{lstlisting}
now reducing = add_reducing nat_rect [lazy, lazy, lazy, eager]
\end{lstlisting}
To get computation going, we need the computation rules for the eliminator:
\begin{lstlisting}
constant nat_β_O :
  ∏ (P : nat → Type) (x : P O) (f : ∏ (n : nat), P n → P (S n)),
    nat_rect P x f O ≡ x

constant nat_β_S :
  ∏ (P : nat → Type) (x : P O) (f : ∏ (n : nat), P n → P (S n))
    (m : nat),
    nat_rect P x f (S m) ≡ f m (nat_rect P x f m)
\end{lstlisting}
which we install as $\beta$-hints:
\begin{lstlisting}
now betas = add_betas [nat_β_O, nat_β_S]
\end{lstlisting}
At this point, we can compute with the recursor, but there is a better way. In Andromeda
there is no built-in notion of ``definition'' at the level of type theory (one can always
use ML-level \lstinline{let}-bindings, but those are always evaluated which has the
undesirable effect of complete unfolding of all definitions). Instead, we break down a
definition into a declaration of the constant and its defining equality. If we install
the defining equality as a $\beta$-hint, a definition behaves like it would in other proof
assistants, but that is just one possibility.

For example, we may define addition as follows:
\begin{lstlisting}
constant ( + ) : nat → nat → nat
constant plus_def :
  ∏ (n m : nat), n + m ≡ nat_rect (λ _, nat) n (λ _ x, S x) m
\end{lstlisting}
Note that \lstinline{plus_def} could be written as
\begin{lstlisting}
constant plus_def' :
  ( + ) ≡ (λ (n m : nat), nat_rect (λ _, nat) n (λ _ x, S x) m)
\end{lstlisting}
The difference between the two is visible when we use them as $\beta$-hints:
\lstinline{plus_def} will unfold only after it has been applied to two arguments, whereas
\lstinline{plus_def'} will do so immediately.

We can derive Peano axioms by using \lstinline{plus_def} as a local $\beta$-hint:
\begin{lstlisting}
let plus_O =
  now betas = add_beta plus_def in
    (λ n, refl n) : ∏ (n : nat), n + O ≡ n

let plus_S =
  now betas = add_beta plus_def in
    (λ n m, refl (n + (S m))) : ∏ (n m : nat), n + (S m) ≡ S (n + m)
\end{lstlisting}
We are free to use the Peano axioms for computation rather than \lstinline{plus_def}, so
we install them globally as $\beta$-hints:
\begin{lstlisting}
now betas = add_betas [plus_O, plus_S]
\end{lstlisting}
It should be clear from this that Andromeda is quite flexible, which is good for
experimentation and tight control of how things are done, but is also bad because the user
has to be more specific in what they want. The overall usability of the system depends on
having a good standard library with sensible default settings.

The definition of multiplication and its Peano axioms are derived similarly:
\begin{lstlisting}
constant ( * ) : nat → nat → nat
constant mult_def :
  ∏ (n m : nat), n * m ≡ nat_rect (λ _, nat) O (λ _ x, x + n) m

let mult_O =
  now betas = add_beta mult_def in
    (λ n, refl O) : ∏ (n : nat), n * O ≡ O

let mult_S =
  now betas = add_beta mult_def in
    (λ n m, refl (n * (S m))) : ∏ (n m : nat), n * (S m) ≡ n * m + n

now betas = add_betas [mult_O, mult_S]
\end{lstlisting}
To compute with numbers, we use the standard library function \lstinline{whnf} that
computes evidence that the given term is equal to its weak head-normal form:
\begin{lstlisting}
do now reducing = add_reducing S [eager] in
   now reducing = add_reducing ( * ) [eager, eager] in
   now reducing = add_reducing ( + ) [eager, eager] in
     whnf ((S (S (S O))) * (S (S (S (S O)))))
\end{lstlisting}
The \lstinline{do} command is the top-level command for evaluating a computation. Notice
that we locally set the arguments of the successor constructor, addition, and
multiplication to be computed eagerly. The effect of this is that the weak head-normal
form is not weak or head-normal anymore, but rather a strongly normalizing call-by-value
strategy. Thus Andromeda outputs
\begin{lstlisting}
⊢ refl (S (S (S (S (S (S (S (S (S (S (S (S O))))))))))))
  : S (S (S O)) * S (S (S (S O))) ≡
    S (S (S (S (S (S (S (S (S (S (S (S O)))))))))))
\end{lstlisting}
It would be easy to obtain just the result, which is the left-hand side of the equality
type. Notice that the proof of equality between $3 \times 4$ and $12$ is a reflexivity
term, even though the normalization procedure generated the proof by stringing together a
large number of reduction steps. In order to keep equality proofs small, the standard
library aggressively replaces equality proofs with reflection terms, using the fact that
whenever $p : \JuEqual{\AA}{\sss}{\ttt}$ then also
$\juRefl{\AA} \ttt : \JuEqual{\AA}{\sss}{\ttt}$.

\subsection{Untyped $\lambda$-calculus}
\label{sec:untyp-lambda-calc}

An example that cannot be done easily in proof assistants based on intensional type
theory is in order. Let us axiomatize the untyped $\lambda$-calculus as a type that is
judgmentally equal to its own function space, and show that it possesses a fixed-point
operator.

We first postulate that there is a type equal to its function space:
\begin{lstlisting}
constant D : Type
constant D_reflexive : D ≡ (D → D)
\end{lstlisting}
We must \emph{not} install \lstinline{D_reflexive} as a $\beta$-hint because it would lead
to non-termination. Instead, we install it and its symmetric version as general hints:
\begin{lstlisting}
now hints = add_hints [D_reflexive, symmetry D_reflexive]
\end{lstlisting}
With these, whenever AML needs to know that \lstinline{D} and \lstinline{D → D} are equal,
the standard library will provide \lstinline{D_reflexive}, or its symmetric version, as
evidence.

Now, we can simply define the fixed-point operator:
\begin{lstlisting}
let fix =
  (λ f,
    let y = (λ x : D, f ((x : D → D) x)) in
    y y)
  : (D → D) → D
\end{lstlisting}
The self-application of \lstinline{x} is well-typed because Andromeda knows that
\lstinline{x} of type \lstinline{D} also has type \lstinline{D → D}, thanks to the hints.
We did have to explicitly coerce \lstinline{x} to the function type. (An alternative would
be to use the coercion mechanism, which is demonstrated in \S\ref{sec:universes}.) Once we overcome the problem of typing the fixed-point
operator, the usual mechanisms suffice to show that it does in fact compute fixed points:
\begin{lstlisting}
let fix_eq =
  (λ f, refl (fix f)) : Π (f : D → D), fix f ≡ f (fix f)
\end{lstlisting}
It is a bit trickier to give a type to a term without weak head-normal form, such as
$(\lambda x.\ x\, x) (\lambda x.\ x\,x)$. We must block $\beta$-reduction of this
particular $\beta$-redex, without blocking all of them. To achieve this, we first
introduce an alias \lstinline{D'} for the type \lstinline{D}:
\begin{lstlisting}
constant D' : Type
constant eq_D_D' : D ≡ D'
\end{lstlisting}
Next, we define the auxiliary term \lstinline{δ} and give it the
type \lstinline{D → D}:
\begin{lstlisting}
let δ = (λ x : D, (x : D → D) x)
\end{lstlisting}
We now form the self-application \lstinline{δ δ} at type \lstinline{D'}:
\begin{lstlisting}
let Ω =
  now hints = add_hints [eq_D_D', symmetry eq_D_D'] in
  (δ : D' → D') (δ : D) : D
\end{lstlisting}
We have the desired term in which $\beta$-reduction is blocked because the inner
$\lambda$-abstractions are typed at \lstinline{D} and the outer application at
\lstinline{D'}. From here, Andromeda happily computes with \lstinline{Ω} without ever
attempting to reduce it (installing \lstinline{eq_D_D'} as a global hint would be a
mistake).

The preceding example should be taken as a proof of concept only. We have reached the
limits of our small standard library. A more serious development of the untyped
$\lambda$-calculus would use a custom equality-checking algorithm instead of manually
juggling hints and type ascriptions.

\subsection{Universes}
\label{sec:universes}

The final example shows how to use coercions and operations to implement a universe à la
Tarski. We postulate a universe \lstinline{U}, whose elements should be thought of as
\emph{names} of types, with an operation \lstinline{El} that converts the names to the
corresponding types:
\begin{lstlisting}
constant U : Type
constant El : U → Type
\end{lstlisting}
Because \lstinline{El} is an eliminator, its normal form should have the argument in
normal form, so we tell the library to normalize it eagerly:
\begin{lstlisting}
now reducing = add_reducing El [eager]
\end{lstlisting}
Next, we postulate that the universe contains names for products and equality types, and
install the relevant equations as $\beta$-hints:
\begin{lstlisting}
constant pi : ∏ (a : U), (El a → U) → U
constant El_pi :
  ∏ (a : U) (b : El a → U), El (pi a b) ≡ (∏ (x : El a), El (b x))
now betas = add_beta El_pi

constant eq : ∏ (a : U), El a → El a → U
constant El_eq :
  ∏ (a : U) (x y : El a), El (eq a x y) ≡ (x ≡ y)
now betas = add_beta El_eq
\end{lstlisting}
For testing purposes we put the name \lstinline{b} of a basic type \lstinline{B} into the
universe:
\begin{lstlisting}
constant B : Type
constant b : U
constant El_b : El b ≡ B
now betas = add_beta El_b
\end{lstlisting}
In principle we can work with \lstinline{U} and \lstinline{El}, but explicit uses of
\lstinline{El} gets tedious quickly. Ideally we want Andromeda to translate between names
and their types automatically, which is achieved with a handler that intercepts coercion
requests. It is easy to coerce names to types with \lstinline{El}, for instance:
\begin{lstlisting}
handle
  (λ x : b, x) : pi b (λ _, b)
with
 | coerce (⊢ ?t : U) (⊢ Type) ⇒ yield (Coercible (El t))
end
\end{lstlisting}
In the $\lambda$-abstraction AML found the name \lstinline{b} but expected a type,
therefore it triggered a coercion operation. The handler intercepted it and yielded
\lstinline{El b}. The process was repeated when AML found \lstinline{pi b (λ _, b)}
instead of a type. The final result printed by Andromeda is
\begin{lstlisting}
⊢ λ (x : El b), x : El (pi b (λ (_ : El b), b))
\end{lstlisting}
We have to work harder to perform the reverse coercion, when a type is encountered where
its code was expected. One first has to implement an AML function \lstinline{name_of} that
takes a type and returns its name, if it can find one. We do not show its
implementation here, and ask the interested readers to consult the examples that come with
the source code. Using \lstinline{name_of} we can handle translation between types and
names in both directions with the handler
\begin{lstlisting}
let universe_handler =
handler
  | coerce (⊢ ?a : U) (⊢ Type) ⇒ yield (Coercible (El a))
  | coerce (⊢ ?T : Type) (⊢ U) ⇒
    match name_of T with
    | None ⇒ yield NotCoercible
    | Some ?name ⇒ yield (Coercible name)
    end
  end
\end{lstlisting}
We added a clause that intercepts coercions from \lstinline{Type} to \lstinline{U}
and uses \lstinline{name_of}. The handler automatically translates names to types and vice
versa. For instance, the computation
\begin{lstlisting}
with universe_handler handle
  (∏ (x : b), x ≡ x) : U
\end{lstlisting}
evaluates to
\begin{lstlisting}
⊢ pi b (λ (y : El b), eq b y y) : U
\end{lstlisting}
In one direction the handler coerced the name \lstinline{b} to the type \lstinline{B}, and in the other the type
\lstinline{∏ (x : B), x ≡ x}
to its name, as shown above.

%%% Local Variables:
%%% mode: latex
%%% TeX-master: "main"
%%% End:

\section{Related work}
\label{sec:related-work}

Andromeda draws heavily on the experience and ideas from other proof assistants. It is
difficult to do justice to all of them. Its overall design follows the tradition of
LCF~\cite{lcf} and its
descendants~\cite{hol_inter_theor_prover,isabelle,harrison:_hol_light}. However, LCF and
many of its descendants use Church's \emph{simple} type
theory~\cite{church40:_formul_simpl_theor_types}, whereas Andromeda is based on the
\emph{dependent} type theory of Martin-Löf~\cite{itt}. Consequently, Andromeda cannot
advantageously integrate the ML-level and the object-level types. There is by necessity a
sharp distinction between the statically typed ML-level and the dynamically evaluated
type-theoretic judgments.

It makes sense to compare Andromeda to other proof assistants based on dependent type
theory~\cite{agda,coq,lean}. For instance, the evaluation strategy for judgments in
Andromeda is based on bidirectional type-checking found in dependently-typed assistants.
Andromeda is primarily a special-purpose programming language, whereas Coq, Agda, and Lean
are tools for interactive proof development. The difference in philosophy of design is
visible in the level of control given to the user. Andromeda gives the user full control
of the system, and expects them to implement their own proof development tools, whereas Coq
and Agda provide more of an end-user environment with a rich selection of ready-made
tools. It is interesting to note that recently Coq and Agda have both started
giving the user more control. New versions of Coq allow the use of tactics inside
type-theoretic terms~\cite[\S2.11.2]{coq-manual} and allow fine-tuning of Coq's unification
algorithm~\cite{coq_unification}. Agda even lets the user install new normalization
rules~\cite{abel16:_sprin} that might break the system.

We already mentioned that NuPRL~\cite{nuprl} validates equality reflection by interpreting
types as partial equivalence relations on terms of a computational model, namely an
extension of the untyped $\lambda$-calculus. We do not wish to make such a commitment in
Andromeda, and instead allow interpretations that are inconsistent with computational type
theory.

%%% Local Variables:
%%% mode: latex
%%% TeX-master: "main"
%%% End:

\section{Future work}
\label{sec:future-work}
\label{sec:remove-type-in-type} % MOVE THIS TO THE MORE SPECIFIC SECTION IF IT IS NUMBERED

We feel that Andromeda shows a promising way to design a proof assistant based on
type theory with equality reflection, but much remains to be done.

\paragraph*{Syntactic sugar and end-user support}

AML turned out to be a useful tool for the implementers of the standard library. If we
imagine that the end-user is a mathematician who just wants to do mathematics, without
learning the intricacies of operations and handlers, then we need further support for
creating a more user friendly environment. There ought to be ways of introducing new
syntactic constructs, and reasonable error reporting by the standard library. We are not
quite sure how to provide such functionality. The approach taken
by Bowman~\cite{bowman16:_growin_proof_assis}
seems interesting. Another possibility is to allow user-defined notations in the style of
Coq, or to completely separate the end-user interface and {AML}.

\paragraph*{Formal verification of the meta-theoretic properties}

While we carefully designed the underlying type theory and made sure it is precisely clear
what the type-theoretic rules are, we have not formally verified that the system has the
desired meta-theoretic properties, such as uniqueness of typing, validity of inversion
principles, and well-behaved context joins. We expect no trouble here, but do insist on
formal verification. In the early stages of implementation we managed to delude ourselves
more than once about the properties of the underlying type theory.

\paragraph*{Recording derivations}

Like all LCF-style proof assistants, Andromeda does not record derivations, only their
conclusions. (In fact, all practical proof assistants do this, though some implement
type theories that allow derivations to be reconstructed.) There
might be situations in which we wish to record or communicate the derivation. For
instance, we might want to send the derivation to another proof assistant for independent
verification. This can be accomplished with a minor modification of {AML}: if we implement
\emph{all} calls of AML to the nucleus as operations (whose default handler is the
nucleus), then the user can intercept them and do whatever they like: record them,
communicate them, or modify them to obtain a proof translation.
Alternatively, the \lstinline|judgment| type in the nucleus could be made into the type of derivations with no breaking changes to the interface.

\paragraph*{Removal of $\Type : \Type$}

A major forthcoming modification of the current system is elimination of $\Type
: \Type$.  The syntax of AML takes advantage of $\Type : \Type$ to conflate term
and type judgments within a single abstract type~$\judgTy$.

But the nucleus does not rely on $\Type : \Type$ at all and separates the various judgment
forms into separate abstract datatypes. There is no technical difficulty in removing
$\Type : \Type$, but the question is what to replace it with. One possibility is to add a
basic type $\mathsf{U}$ and a basic type family $\mathsf{El}$ indexed by $\mathsf{U}$, and
then use these as a Tarski-style universe, with a standard library employing techniques of
\S\ref{sec:universes} to make the system usable. (This can be extended to multiple
universes if desired.) Paolo Capriotti's recently suggested such a
setup~\cite{capriotti17:_notion}, based on semantic considerations in categories of
presheaves. Finally, AML could be modified to support several abstract datatypes, one for
each form of object-level judgment.

Once this is done, several interesting possibilities arise. The user could hypothesize any
universe structure they like, including putting back $\Type : \Type$. We might even be
able to remove equality reflection from the nucleus, and make it user-definable. We hope
to report on these exciting developments in the near future.

%%% Local Variables:
%%% mode: latex
%%% TeX-master: "main"
%%% End:

% \subparagraph*{Acknowledgments.}
% We thank someone.

%%
%% Bibliography
%%

\bibliographystyle{plain}
\bibliography{andromeda}

\appendix

\section{The rules of type theory}
\label{sec:rules-type-theory}

% Used only here to delimit groups of rules (\paragraph seems too bold)
\newcommand{\explanation}[1]{\goodbreak\par\noindent #1}

In this appendix we give the formulation of type theory in a declarative way that
minimizes the number of judgments, and so is better suited for a semantic account. We omit formal
treatment of bound variables and substitution, which is standard.

\subsection{Syntax}
\label{sec:syntax}

\begin{align*}
  \intertext{Contexts}
  \G, \D
    \bnf   {}& \ctxempty & & \text{empty context}\\
    \bnfor {}& \ctxextend{\G}{\x}{\AA} & & \text{context $\G$ extended with $\x : \AA$}
  \intertext{Terms and types}
  \sss, \ttt, \AA, \BB
    \bnf   {}& \Type & & \text{universe}\\
    \bnfor {}& \Prod{x}{\AA} \BB & & \text{product}\\
    \bnfor {}& \JuEqual{\AA}{\sss}{\ttt} & & \text{equality type} \\
    \bnfor {}&  \x   &&\text{variable} \\
    \bnfor {}&  \lam{\x}{\AA}{\BB} \ttt  &&\text{$\lambda$-abstraction} \\
    \bnfor {}&  \app{\sss}{\x}{\AA}{\BB}{\ttt}  &&\text{application} \\
    \bnfor {}&  \juRefl{\AA} \ttt  &&\text{reflexivity}
\end{align*}

\subsection{Judgments}
\label{sec:judgments}

\begin{align*}
& \isctx{\G}                   & & \text{$\G$ is a well formed context} \\
& \isterm{\G}{\ttt}{\AA}       & & \text{$\ttt$ is a well formed term of type $\AA$ in context $\G$} \\
& \eqterm{\G}{\sss}{\ttt}{\AA} & & \text{$\sss$ and $\ttt$ are equal terms of type $\AA$ in context $\G$}
\intertext{We use the following abbreviations:}
& \istype{\G}{\AA}             & & \text{abbreviates $\isterm{\G}{\AA}{\Type}$} \\
& \eqtype{\G}{\AA}{\BB}        & & \text{abbreviates $\eqterm{\G}{\AA}{\BB}{\Type}$}
\end{align*}

\subsection{Contexts}
\label{sec:contexts}

\begin{mathpar}
  \infer[\rulename{ctx-empty}]
  { }
  {\isctx{\ctxempty}}

  \infer[\rulename{ctx-extend}]
  {\isctx{\G} \\
   \istype{\G}{\AA} \\
   \x \not\in \ctxdom{\G}
  }
  {\isctx{(\ctxextend{\G}{\x}{\AA})}}
\end{mathpar}

% In rules that extend the context, we leave implicit the premise that the extended context be well formed.
% XXX Is this covered by the term-var rule?

\subsection{Terms and types}

\explanation{Conversion}
\begin{mathpar}
  \infer[\rulename{term-ty-conv}]
  {\isterm{\G}{\ttt}{\AA} \\
   \eqtype{\G}{\AA}{\BB}
  }
  {\isterm{\G}{\ttt}{\BB}}
\end{mathpar}

\explanation{Variable}
\begin{mathpar}
  \infer[\rulename{term-var}]
  {\isctx{(\ctxextend{\G}{\x}{\AA})}}
  {\isterm{\ctxextend{\G}{\x}{\AA}}{\x}{\AA}}

  \infer[\rulename{term-var-skip}]
  {\isctx{(\ctxextend{\G}{\y}{\BB})} \\
   \isterm{\G}{\x}{\AA}
  }
  {\isterm{\ctxextend{\G}{\y}{\BB}}{\x}{\AA}}
\end{mathpar}

\explanation{Universe}
\begin{mathpar}
  \infer[\rulename{ty-type}]
  {\isctx{\G}
  }
  {\istype{\G}{\Type}}
\end{mathpar}

\explanation{Product}
\begin{mathpar}
  \infer[\rulename{ty-prod}]
  {\istype{\G}{\AA} \\
   \istype{\ctxextend{\G}{\x}{\AA}}{\BB}
  }
  {\istype{\G}{\Prod{\x}{\AA}{\BB}}}

  \infer[\rulename{term-abs}]
  {\isterm{\ctxextend{\G}{\x}{\AA}}{\ttt}{\BB}}
  {\isterm{\G}{(\lam{\x}{\AA}{\BB}{\ttt})}{\Prod{\x}{\AA}{\BB}}}

  \infer[\rulename{term-app}]
  {\isterm{\G}{\sss}{\Prod{x}{\AA} \BB} \\
   \isterm{\G}{\ttt}{\AA}
  }
  {\isterm{\G}{\app{\sss}{\x}{\AA}{\BB}{\ttt}}{\subst{\BB}{\x}{\ttt}}}
\end{mathpar}

\explanation{Equality type}
\begin{mathpar}
  \infer[\rulename{ty-eq}]
  {\istype{\G}{\AA}\\
   \isterm{\G}{\sss}{\AA}\\
   \isterm{\G}{\ttt}{\AA}
  }
  {\istype{\G}{\JuEqual{\AA}{\sss}{\ttt}}}

  \infer[\rulename{term-refl}]
  {\isterm{\G}{\ttt}{\AA}}
  {\isterm{\G}{\juRefl{\AA} \ttt}{\JuEqual{\AA}{\ttt}{\ttt}}}
  \end{mathpar}

\subsection{Equality}

\explanation{General rules}
\begin{mathpar}
  \infer[\rulename{eq-refl}]
  {\isterm{\G}{\ttt}{\AA}}
  {\eqterm{\G}{\ttt}{\ttt}{\AA}}

  \infer[\rulename{eq-sym}]
  {\eqterm{\G}{\ttt}{\sss}{\AA}}
  {\eqterm{\G}{\sss}{\ttt}{\AA}}

  \infer[\rulename{eq-trans}]
  {\eqterm{\G}{\sss}{\ttt}{\AA}\\
   \eqterm{\G}{\ttt}{\uuu}{\AA}}
  {\eqterm{\G}{\sss}{\uuu}{\AA}}
\end{mathpar}

\explanation{Conversion}
\begin{mathpar}
  \infer[\rulename{eq-ty-conv}]
  {\eqterm{\G}{\sss}{\ttt}{\AA}\\
    \eqtype{\G}{\AA}{\BB}}
  {\eqterm{\G}{\sss}{\ttt}{\BB}}
\end{mathpar}

\explanation{Equality reflection}
\begin{mathpar}
  \infer[\rulename{eq-reflection}]
  {\isterm{\G}{\uuu}{\JuEqual{\AA}{\sss}{\ttt}}}
  {\eqterm{\G}{\sss}{\ttt}{\AA}}
\end{mathpar}

\explanation{Computation}
\begin{mathpar}
  \infer[\rulename{prod-beta}]
  {
    \isterm{\ctxextend{\G}{\x}{\AA}}{\sss}{\BB}\\
    \isterm{\G}{\ttt}{\AA}}
  {\eqterm{\G}{\app{(\lam{\x}{\AA}{\BB}{\sss})}{\x}{\AA}{\BB}{\ttt}}
              {\subst{\sss}{\x}{\ttt}}
              {\subst{\BB}{\x}{\ttt}}}
\end{mathpar}

\explanation{Extensionality}
\begin{mathpar}
  \infer[\rulename{eq-eta}]
  {\isterm{\G}{\ttt}{\JuEqual{\AA}{\sss}{\uuu}} \\
    \isterm{\G}{\vvv}{\JuEqual{\AA}{\sss}{\uuu}}
  }
  {\eqterm{\G}{\ttt}{\vvv}{\JuEqual{\AA}{\sss}{\uuu}}}

  \infer[\rulename{prod-eta}]
  {\isterm{\G}{\sss}{\Prod{\x}{\AA}{\BB}}\\
   \isterm{\G}{\ttt}{\Prod{\x}{\AA}{\BB}}\\\\
   \eqterm{\ctxextend{\G}{\x}{\AA}}{(\app{\sss}{\x}{\AA}{\BB}{\x})}
          {(\app{\ttt}{\x}{\AA}{\BB}{\x})}{\BB}
  }
  {\eqterm{\G}{\sss}{\ttt}{\Prod{\x}{\AA}{\BB}}}
\end{mathpar}

\subsubsection{Congruences}

\explanation{Type formers}
\begin{mathpar}
  \infer[\rulename{cong-prod}]
  {\eqtype{\G}{\AA}{\CC}\\
   \eqtype{\ctxextend{\G}{\x}{\AA}}{\BB}{\subst{\DD}{\y}{\x}}}
  {\eqtype{\G}{\Prod{\x}{\AA}{\BB}}{\Prod{\y}{\CC}{\DD}}}

  \infer[\rulename{cong-eq}]
  {\eqtype{\G}{\AA}{\BB}\\
   \eqterm{\G}{\sss}{\uuu}{\AA}\\
   \eqterm{\G}{\ttt}{\vvv}{\AA}
  }
  {\eqtype{\G}{\JuEqual{\AA}{\sss}{\ttt}}
              {\JuEqual{\BB}{\uuu}{\vvv}}}
\end{mathpar}

\explanation{Products}
\begin{mathpar}
  \infer[\rulename{cong-abs}]
  {\eqtype{\G}{\AA}{\CC}\\
    \eqtype{\ctxextend{\G}{\x}{\AA}}{\BB}{\subst{\DD}{\y}{\x}}\\
    \eqterm{\ctxextend{\G}{\x}{\AA}}{\sss}{\subst{\ttt}{\y}{\x}}{\BB}}
  {\eqterm{\G}{(\lam{\x}{\AA}{\BB}{\sss})}
              {(\lam{\y}{\CC}{\DD}{\ttt})}
              {\Prod{\x}{\AA}{\BB}}}

  \infer[\rulename{cong-app}]
  {\eqtype{\G}{\AA}{\CC}\\
   \eqtype{\ctxextend{\G}{\x}{\AA}}{\BB}{\subst{\DD}{\y}{\x}}\\\\
   \eqterm{\G}{\sss}{\uuu}{\Prod{\x}{\AA}{\BB}}\\
   \eqterm{\G}{\ttt}{\vvv}{\AA}}
  {\eqterm{\G}{(\app{\sss}{\x}{\AA}{\BB}{\ttt})}{(\app{\uuu}{\y}{\CC}{\DD}{\vvv})}{\subst{\BB}{\x}{\ttt}}}
\end{mathpar}

\explanation{Equality types}
\begin{mathpar}
\infer[\rulename{cong-refl}]
{\eqtype{\G}{\AA}{\BB}\\
 \eqterm{\G}{\sss}{\ttt}{\AA}
}
{\eqterm{\G}{\juRefl{\AA} \sss}{\juRefl{\BB} \ttt}{\JuEqual{\AA}{\sss}{\sss}}}
\end{mathpar}

%%% Local Variables:
%%% mode: latex
%%% TeX-master: "main"
%%% End:

\section{The \lstinline{auto} tactic}
\label{sec:auto-tactic}

We include here the complete code for implementing a simple \lstinline{auto} tactic
from~\S~\ref{sec:system-design}.

We first define the \lstinline{map} function to show how AML syntax works, and the
auxiliary \lstinline{apply} function that folds application of a function over a list of
arguments:
\begin{lstlisting}
let rec map f xs =
  match xs with
  | [] ⇒ []
  | ?x :: ?xs ⇒ (f x) :: (map f xs)
  end

let rec apply f xs =
  match xs with
  | [] ⇒ f
  | ?x :: ?xs ⇒ apply (f x) xs
  end
\end{lstlisting}
Next, we declare the \lstinline{failure} operation that is triggered when the search fails:
\begin{lstlisting}
operation failure : judgment
\end{lstlisting}
Next we define the function \lstinline{subgoals} that takes a goal \lstinline{A} and a
hypothesis \lstinline{B} and computes a list of subgoals that together with \lstinline{B}
imply \lstinline{A}. For instance, if \lstinline{B} is equal to
\lstinline{C → D → A}
then the computed subgoals are \lstinline{[C, D]}:
\begin{lstlisting}
let rec subgoals A B =
  match B with
  | ⊢ A ⇒ []
  | ⊢ ?P → ?Q ⇒ P :: (subgoals A Q)
  | _ ⇒ [failure]
  end
\end{lstlisting}
The function \lstinline{derive} takes a goal \lstinline{A} and attempts to derive it. If
the goal is an implication, it introduces the antecedent as a hypothesis and calls itself
recursively. Otherwise it tries to prove the goal from the current hypotheses by simple
backchaining:
\begin{lstlisting}
let rec derive A =
  match A with
  | ⊢ ?P → ?Q ⇒ λ (x : P), derive Q
  | ⊢ _ ⇒ backchain A hypotheses
  end

and backchain A lst =
  match lst with
  | [] ⇒ failure
  | (⊢ ?f : ?B) :: ?lst ⇒
    handle
      apply f (map derive (subgoals A B))
    with
      failure ⇒ backchain A lst
    end
  end
\end{lstlisting}
Note how \lstinline{backchain} uses a handler to intercept \lstinline{failure}, just like
an ordinary exception handler does. Finally, we declare an operation \lstinline{auto} and
define a global handler that handles it. The handler only works when \lstinline{auto} is
used in checking mode:
\begin{lstlisting}
operation auto : judgment

handle
| auto : ?T' ⇒
    match T' with
    | Some ?T ⇒ derive T
    | None ⇒ failure
    end
end
\end{lstlisting}
Now we can use \lstinline{auto} to inhabit types. For example,
\begin{lstlisting}
(λ (X : Type), auto : X → X)
\end{lstlisting}
computes to
\begin{lstlisting}
⊢ λ (X : Type) (x : X), x : Π (X : Type), X → X
\end{lstlisting}
and given the types
\begin{lstlisting}
constant A : Type
constant B : Type
constant C : Type
\end{lstlisting}
the computation
\begin{lstlisting}
auto : (A → B → C) → (A → B) → (A → C)
\end{lstlisting}
results in
\begin{lstlisting}
⊢ λ (x : A → B → C) (x0 : A → B) (x1 : A), x x1 (x0 x1)
  : (A → B → C) → (A → B) → A → C
\end{lstlisting}

%%% Local Variables:
%%% mode: latex
%%% TeX-master: "main"
%%% End:

\end{document}